\documentclass[10pt,onecolumn,showpacs,showkeys,amsmath,amssymb,aps,pra,superscriptaddress,notitlepage,floatfix]{revtex4-2}
\usepackage[utf8]{inputenc}
\usepackage{mathtools}
\usepackage{bm}
\usepackage{amsmath}
\usepackage{amsthm}
\usepackage{graphicx}

\makeatletter

\usepackage{stylesetting}
\usepackage{booktabs}
\usepackage{makecell}

\makeatother

\usepackage[hidelinks,linktoc=all,bookmarksnumbered]{hyperref}

\makeatletter
\newcommand{\appendixcontents}{%
  \begingroup
  \renewcommand{\tocname}{Appendix Contents}%
  \let\appendix@savedcontentsline\contentsline
  \renewcommand{\contentsline}[4]{}%
  \let\appendix@savedtoc\appendix@toc
  \def\appendix@toc{%
    \let\contentsline\appendix@savedcontentsline
    \appendix@savedtoc
  }%
  \tableofcontents
  \endgroup
}
\makeatother

\begin{document}
\setlength{\emergencystretch}{2em}
\title{Distributed Inner-Product
Estimation for Continuous-Variable Quantum States}
\date{October 29, 2026}

\author{Jun Tang}
\affiliation{National Mobile Communications Research Laboratory, Southeast University,
Nanjing 210096, China}
\affiliation{Frontiers Science Center for Mobile Information Communication and
Security, Southeast University, Nanjing 210096, China}
\author{Congcong Zheng}
\affiliation{State Key Lab of Millimeter Waves, Southeast University, Nanjing 211189,
China}
\affiliation{Purple Mountain Laboratories, Nanjing 211111, China}
\affiliation{Frontiers Science Center for Mobile Information Communication and
Security, Southeast University, Nanjing 210096, China}
\author{Xutao Yu}
\affiliation{State Key Lab of Millimeter Waves, Southeast University, Nanjing 211189,
China}
\affiliation{Purple Mountain Laboratories, Nanjing 211111, China}
\affiliation{Frontiers Science Center for Mobile Information Communication and
Security, Southeast University, Nanjing 210096, China}
\author{Zaichen Zhang}
\thanks{Corresponding author: zczhang@seu.edu.cn}
\affiliation{National Mobile Communications Research Laboratory, Southeast University,
Nanjing 210096, China}
\affiliation{Purple Mountain Laboratories, Nanjing 211111, China}
\affiliation{Frontiers Science Center for Mobile Information Communication and
Security, Southeast University, Nanjing 210096, China}
\begin{abstract}
Distributed inner product estimation (DIPE) is a fundamental task in quantum information, aiming to estimate the inner product between two unknown quantum states prepared on distributed quantum platforms. 
For multimode continuous-variable (CV) systems, the infinite-dimensional Hilbert space poses an additional challenge. 
In this work, we develop a CV-DIPE protocol using only local homodyne measurements at shared phase settings and classical communication. 
Under a known bound on the mean energy per mode, we derive explicit finite-sample guarantees and sufficient copy counts for estimating the full inner product to prescribed accuracy and confidence. 
The energy constraint yields an inverse-square truncation-bias bound, while Gaussian structure strengthens the tail control to exponential decay and improves the resulting accuracy scaling. 
Numerical simulations with Gaussian and non-Gaussian states illustrate the balance between truncation bias and statistical error and its dependence on mode number and energy.
Our results enable independent CV platforms to benchmark their state preparations against one another, even when neither state admits a tractable classical description.
\end{abstract}

\maketitle

\section{Introduction}

Bosonic continuous-variable (CV) systems provide a versatile setting for
quantum information processing~\cite{BraunsteinVanLoock2005CV,Andersen2010CVProcessing}, with information encoded in the
quadrature degrees of freedom of optical and microwave
fields~\cite{Blais2021CircuitQED,Mallet2011MicrowaveTomography}.
Gaussian operations and quadrature measurements provide well-established
tools for state preparation, manipulation, and characterization~\cite{Weedbrook2012GQI,Larsen2021MultimodeGates}, while
suitable non-Gaussian resources~\cite{Andersen2015Hybrid,Walschaers2021NonGaussian} extend these capabilities to universal
quantum computation~\cite{LloydBraunstein1999CVQC,Menicucci2006CVCluster}.
Advances in multimode entanglement generation~\cite{Yokoyama2013LargeCluster,Asavanant2019Cluster,Larsen2019Cluster,Jia2025IntegratedMultimode}, integrated quantum
photonics~\cite{Lenzini2018IntegratedCV,Arrazola2021ProgrammableChip,Clark2026IntegratedCV}, and quantum networks~\cite{Wehner2018QuantumInternet,Guo2020CVNetwork,AghaeeRad2025ModularPhotonics} are extending CV quantum information
processing from individual devices to increasingly complex
systems.
As independently operated platforms become capable of preparing
increasingly complex quantum states, a fundamental question arises:
when neither state admits a tractable classical description and joint
quantum measurements across platforms are unavailable, how can their
preparations be quantitatively compared using local operations and classical communication (LOCC)?

This task is commonly referred to as cross-platform comparison, with
distributed inner-product estimation (DIPE) as a central
primitive~\cite{Elben2020CrossPlatform,Anshu2022DIPE,Zhu2022CrossPlatform,Knorzer2023Networks,Zheng2024CrossPlatformProcesses,Hinsche2025DIPE,Zheng2025ChannelSimilarity,Gong2026PurityInnerProduct,Zheng2026StructuredDIPE,Dalton2025ModularVerification,Fanizza2020Overlap}.
DIPE aims to estimate the inner product
$f(\rho,\sigma):=\operatorname{Tr}(\rho\sigma)$ between two unknown
quantum states using only LOCC.
These studies have primarily focused on finite-dimensional,
discrete-variable systems, raising the question of how to realize
DIPE for multimode CV states using experimentally accessible measurements.
Realizing this task in CV systems requires connecting the inner product to
experimentally accessible measurement data. Homodyne detection has long
been a standard tool for CV quantum-state
tomography~\cite{Smithey1993HomodyneTomography,Breitenbach1997QuantumStates,LvovskyRaymer2009CVTomography} and has more recently been
incorporated into classical-shadow protocols for estimating selected
state properties~\cite{GandhariEtAl2024,BeckerEtAl2024}.
Related work has also investigated the estimation of
characteristic-function values at prescribed phase-space points under
suitable assumptions on the state~\cite{WuZhuChiribellaLiu2024CVLearning}.
In particular, Fourier transforms of the quadrature outcome
distributions obtained by homodyne detection determine the
characteristic function on the corresponding phase-space
slices~\cite{VogelRisken1989,LvovskyRaymer2009CVTomography}.
Thus, homodyne data provide access not only to full state reconstruction
but also to the direct estimation of selected state properties.
Since the inner product admits a phase-space integral representation
in terms of the two states' characteristic functions~\cite{Werner1984},
this connection suggests a natural question: can homodyne data acquired
separately on two platforms be used to estimate the inner product between
unknown states without reconstructing either state?
Recent nonparametric kernel methods have already used quadrature data
to estimate inner products and other trace functionals, providing further
support for this approach~\cite{markovich2026}.
These advances motivate us to investigate DIPE protocols for general
multimode CV states, with rigorous guarantees on estimation error and
measurement resources under explicit physical assumptions.

In this paper, we develop a homodyne-only protocol for estimating the
inner product between two independently prepared, unknown multimode CV
states and establish rigorous finite-sample guarantees
for the full inner product under the stated assumptions on the states. For fixed mode number $M$, per-mode mean-energy
bound $E$, and number $n$ of homodyne shots per state at each phase setting,
Hoeffding's inequality~\cite{Hoeffding1963} yields sufficient total numbers of state copies scaling as $O\!\bigl(\epsilon^{-(2M+2)}\log(2/\delta)\bigr)$ for general
energy-constrained states and
$O\!\bigl(\epsilon^{-2}[\log(2M/\epsilon)]^{2M}\log(2/\delta)\bigr)$ for
Gaussian states satisfying the same energy constraint, ensuring error at
most $\epsilon$ with confidence at least $1-\delta$. Thus, in these
sufficient bounds, Gaussian structure replaces the additional accuracy
cost $\epsilon^{-2M}$ with the polylogarithmic factor
$[\log(2M/\epsilon)]^{2M}$. A complementary Bernstein
bound~\cite{Boucheron2013Concentration} explicitly incorporates the
dependence on the number of shots per phase setting through the
single-setting variance upper bound.

This paper is organized as follows.
Sec.~\ref{sec:problem_setting} introduces preliminaries.
Sec.~\ref{sec:truncated_estimator} presents the estimator protocol.
Sec.~\ref{sec:copy_complexity} derives sample-complexity bounds under different assumptions on the states.
Sec.~\ref{sec:numerics_plan} evaluates the estimator numerically for
representative CV state families. Finally, we conclude in
Sec.~\ref{sec:conclusions}.

\section{HOMODYNE ACCESS TO STATE OVERLAP}
\label{sec:problem_setting}

Consider an $M$-mode bosonic system with Hilbert space
$\mathcal{H}_{M}=\bigotimes_{\ell=1}^{M}\mathcal{H}_{\ell}$, and set
$\hbar=1$. For each mode $\ell$ and $m$, the annihilation and creation operators
satisfy $[a_{\ell},a_{m}^{\dagger}]=\delta_{\ell m}$, and the canonical
quadratures are
$\hat q_{\ell}=(a_{\ell}+a_{\ell}^{\dagger})/\sqrt{2}$ and
$\hat p_{\ell}=(a_{\ell}-a_{\ell}^{\dagger})/(i\sqrt{2})$.
We collect them as
$\hat{\mathbf{R}}=(\hat q_{1},\hat p_{1},\ldots,\hat q_{M},\hat p_{M})^{\mathsf T}$,
with $[\hat R_j,\hat R_k]=i\Omega_{jk}$ and
$\Omega=\bigoplus_{\ell=1}^{M}
\begin{psmallmatrix}0&1\\-1&0\end{psmallmatrix}$.
For $\mathbf{u}\in\mathbb{R}^{2M}$, define the Weyl operator
$W(\mathbf{u}):=\exp(i\mathbf{u}^{\mathsf T}\hat{\mathbf{R}})$ and the characteristic function
$\chi_{\tau}(\mathbf{u}):=\operatorname{Tr}[\tau W(\mathbf{u})]$.

Let \(\rho\) and \(\sigma\) denote the density operators of the quantum states prepared on two independent quantum platforms. The quantity of interest is their inner product. With the above
convention, the Plancherel identity~\cite{Werner1984} gives
\begin{equation}
f(\rho,\sigma)=\operatorname{Tr}(\rho\sigma)
=
\frac{1}{(2\pi)^{M}}
\int_{\mathbb{R}^{2M}}
\chi_{\rho}(\mathbf{u})\chi_{\sigma}(-\mathbf{u})\,d\mathbf{u}.
\label{eq:phase_space_identity_main}
\end{equation}
Thus, direct inner-product estimation can be formulated as the estimation of
an integral involving the characteristic functions of the two states.

\begin{figure}[t]
\centering
\includegraphics[width=0.60\linewidth]{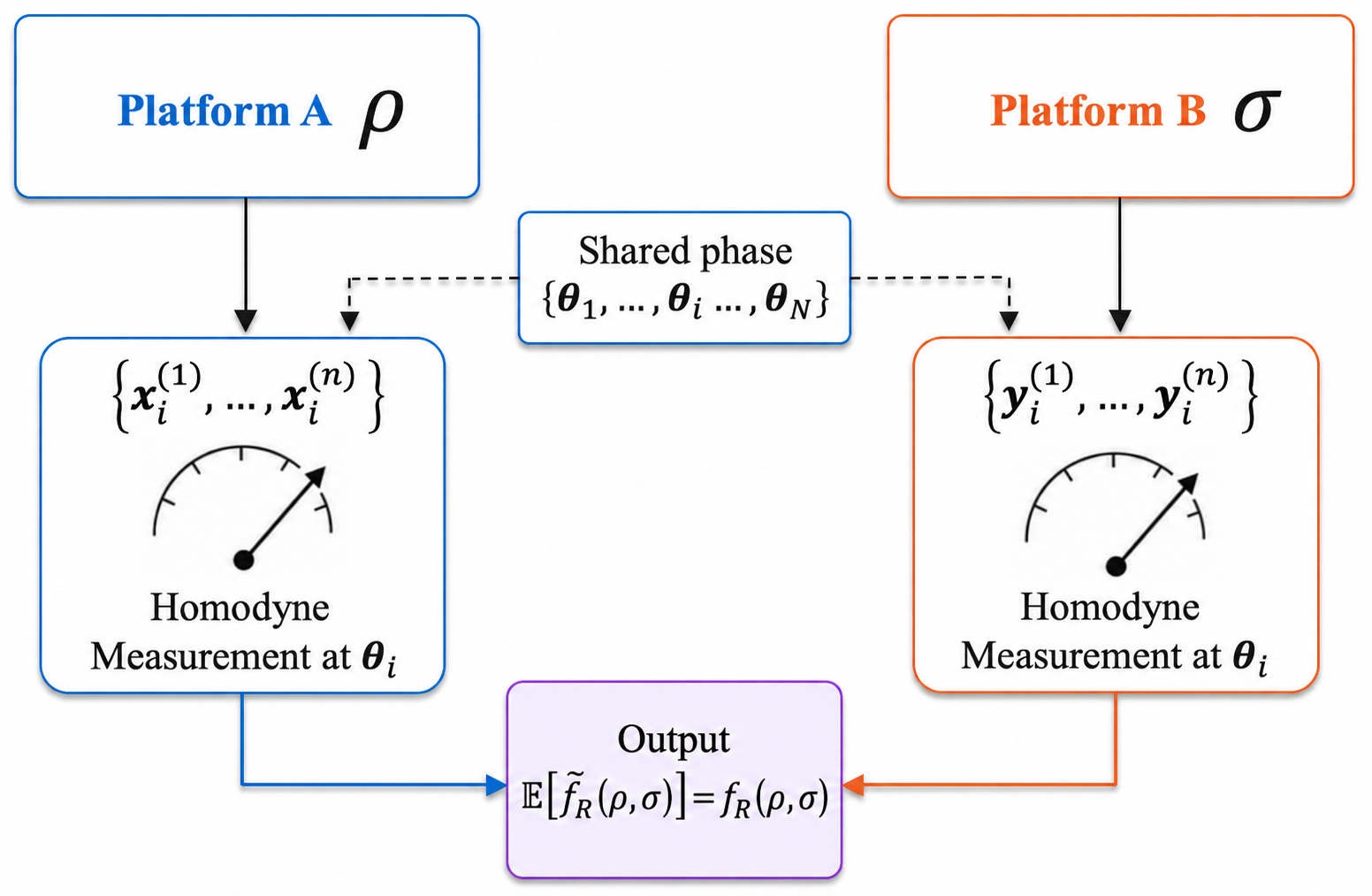}
\caption{Schematic workflow for continuous-variable distributed inner-product estimation (CV-DIPE). Platform A prepares the state described by \(\rho\), and platform B prepares the state described by \(\sigma\). For each shared phase setting \(\boldsymbol{\theta}_i\), \(i=1,\ldots,N_\theta\), each platform performs \(n\) local homodyne measurements, with each measurement producing an \(M\)-dimensional outcome vector. The corresponding records are denoted by \(\{\mathbf{x}_i^{(1)},\ldots,\mathbf{x}_i^{(n)}\}\) and \(\{\mathbf{y}_i^{(1)},\ldots,\mathbf{y}_i^{(n)}\}\).
Classical post-processing of these records yields an unbiased estimator
of the truncated inner product, satisfying
$\mathbb{E}[\widetilde f_R(\rho,\sigma)]=f_R(\rho,\sigma)$. Here,
$R$ is the truncation radius, and the expectation is taken over the randomly
sampled phase settings and homodyne measurement outcomes.}
\label{fig:cross_platform_workflow}
\end{figure}

We assume that the two
platforms use the same quadrature normalization and are calibrated to a
common phase reference, so that the same phase setting
corresponds to the same phase-space directions on both platforms.
Local homodyne detection then enables estimation of the characteristic functions at specified phase-space points in Eq.~\eqref{eq:phase_space_identity_main}.
Fig.~\ref{fig:cross_platform_workflow} summarizes this measurement setup. The two platforms share a sequence of phase
settings, collect conditionally independent local measurement records,
and combine the data only through classical post-processing.

To connect the phase-space integral with homodyne measurement data, write
each mode component as
$(u_{q,\ell},u_{p,\ell})
=s_{\ell}(\cos\theta_{\ell},\sin\theta_{\ell})$,
where $\theta_{\ell}\in[0,\pi)$ and $s_{\ell}\in\mathbb{R}$. Define
\begin{equation}
\boldsymbol{\xi}(\boldsymbol{\theta},\mathbf{s})
:=
\big(
s_{1}\cos\theta_{1},
s_{1}\sin\theta_{1},
\ldots,
s_{M}\cos\theta_{M},
s_{M}\sin\theta_{M}
\big)^{\mathsf T}.
\label{eq:slice_map_main}
\end{equation}
Here, \(\mathbf{s}\) is a classical Fourier variable used only in post-processing. Each component \(s_{\ell}\) may take either sign: \(|s_{\ell}|\) gives the radial coordinate, while the sign selects one of the two opposite phase-space directions under the convention \(\theta_{\ell}\in[0,\pi)\). Every \(\mathbf{u}\in\mathbb{R}^{2M}\) admits such a representation, which is unique almost everywhere.

For phase $\theta_{\ell}$, the homodyne observable of mode $\ell$ is
$\hat X_{\theta_{\ell}}^{(\ell)}
=\hat q_{\ell}\cos\theta_{\ell}
+\hat p_{\ell}\sin\theta_{\ell}$.
A multimode homodyne measurement with phase setting
$\boldsymbol{\theta}=(\theta_{1},\ldots,\theta_{M})$ produces an outcome
vector $\mathbf{x}\in\mathbb{R}^{M}$ distributed according to
$p_{\tau}(\mathbf{x}\mid\boldsymbol{\theta})$.
The standard homodyne Fourier-slice relation gives the following direct connection between homodyne outcomes and
the phase-space integrand.

\begin{lemma}[Homodyne Access to the Phase-Space Inner-Product Integrand~\cite{Smithey1993HomodyneTomography,WuZhuChiribellaLiu2024CVLearning}]
\label{lem:homodyne_access}
For $\mathbf{u}=\boldsymbol{\xi}(\boldsymbol{\theta},\mathbf{s})$,
\begin{equation}
\mathbb{E}_{\mathbf{x}\sim
p_{\tau}(\cdot\mid\boldsymbol{\theta})}
\left[e^{i\mathbf{s}^{\mathsf T}\mathbf{x}}\right]
=
\chi_{\tau}(\mathbf{u}).
\label{eq:fourier_slice_main}
\end{equation}
For conditionally independent outcomes
$\mathbf{x}\sim p_{\rho}(\cdot\mid\boldsymbol{\theta})$ and
$\mathbf{y}\sim p_{\sigma}(\cdot\mid\boldsymbol{\theta})$
obtained at the same phase setting,
\begin{equation}
\mathbb{E}_{\mathbf{x},\mathbf{y}\mid\boldsymbol{\theta}}
\left[
e^{i\mathbf{s}^{\mathsf T}(\mathbf{x}-\mathbf{y})}
\right]
=
\chi_{\rho}(\mathbf{u})\chi_{\sigma}(-\mathbf{u}).
\label{eq:main_accessible_product}
\end{equation}
\end{lemma}
Eq.~\eqref{eq:fourier_slice_main} establishes that the Fourier transform of the homodyne outcome distribution gives the characteristic function on the phase-space slice specified by the phase setting. When the two platforms perform measurements at the same phase setting, the conditional independence of their outcomes given the shared phase setting allows the expectation in Eq.~\eqref{eq:main_accessible_product} to factorize, yielding precisely the product $\chi_\rho(\mathbf u)\chi_\sigma(-\mathbf u)$ appearing in the phase-space representation of the inner product.  This relation
provides the measurement basis for the truncated phase-space estimator
introduced in the next section.

\section{ Estimation Protocol}
\label{sec:truncated_estimator}
\vspace*{0.15cm}
Building on the pointwise homodyne relation above (Lemma~\ref{lem:homodyne_access}), we now construct a finite-sample protocol for CV-DIPE and establish the unbiasedness of the estimator.

\paragraph*{\textbf{Phase-Space Truncation.}}
To obtain bounded statistics for finite-sample error control, we truncate the phase-space integral to a product of modewise disks of radius $R$.

Specifically, for $R>0$, we define the truncation domain as a modewise product of disks,
\begin{equation}
\mathcal{D}_{R}
:=
\prod_{\ell=1}^{M}
\left\{
(u_{q,\ell},u_{p,\ell})\in\mathbb{R}^{2}:
u_{q,\ell}^{2}+u_{p,\ell}^{2}\le R^{2}
\right\}.
\label{eq:main_disk_domain}
\end{equation}
This truncation domain is rotationally symmetric within each single-mode phase
plane, does not select a preferred quadrature direction, and requires no
additional structural assumption on the states. The corresponding
truncated inner product is
\begin{equation}
f_{R}(\rho,\sigma)
:=
\frac{1}{(2\pi)^{M}}
\int_{\mathcal{D}_{R}}
\chi_{\rho}(\mathbf{u})\chi_{\sigma}(-\mathbf{u})\,d^{2M}\mathbf{u}.
\label{eq:main_truncated_target}
\end{equation}
For any quantum state $\tau$, the characteristic function satisfies
$\chi_{\tau}(-\mathbf{u})=\overline{\chi_{\tau}(\mathbf{u})}$.
Since $\mathcal{D}_{R}$ is symmetric under
$\mathbf{u}\mapsto-\mathbf{u}$, the imaginary contributions
cancel upon integration. Thus, the truncated inner product
can be expressed as
\begin{equation}
f_R(\rho,\sigma)
=
\frac{1}{(2\pi)^M}
\int_{\mathcal{D}_R}
\operatorname{Re}\!\left[
\chi_\rho(\mathbf{u})\chi_\sigma(-\mathbf{u})
\right]
\,d^{2M}\mathbf{u}.
\label{eq:main_truncated_target_real}
\end{equation}
\vspace*{0.15cm}
\paragraph*{\textbf{Homodyne Estimator for the Truncated Inner Product.}}

Fix a truncation radius $R>0$ and positive integers $n,N_\theta$.
For each $i=1,\ldots,N_\theta$, independently draw
$\boldsymbol\theta_i\sim\operatorname{Unif}([0,\pi)^M)$ and measure
$n$ independent copies of each state at this shared phase setting.
The outcomes are
$\mathbf x_i^{(a)}\sim p_\rho(\cdot\mid\boldsymbol\theta_i)$ and
$\mathbf y_i^{(b)}\sim p_\sigma(\cdot\mid\boldsymbol\theta_i)$,
for $a,b=1,\ldots,n$, with the two platform samples conditionally
independent given $\boldsymbol\theta_i$. The setting record is
\begin{equation}
\mathsf D_i
:=
\left(
\boldsymbol{\theta}_i,
\{\mathbf{x}_i^{(a)}\}_{a=1}^{n},
\{\mathbf{y}_i^{(b)}\}_{b=1}^{n}
\right).
\label{eq:RB_data_record}
\end{equation}
We write $\mathsf D:=(\mathsf D_1,\ldots,\mathsf D_{N_\theta})$
for the complete data. The measurement records are random and independent across phase settings.

For radial integration, introduce a vector
$\mathbf S_i=(S_{i,1},\ldots,S_{i,M})^{\mathsf T}$ independent of
$\mathsf D_i$, with independent components of density
\begin{equation}
q_R(s)=\frac{|s|}{R^2}\mathrm{1}_{[-R,R]}(s).
\label{eq:signed_coordinate_density}
\end{equation}
Here, $\mathrm{1}_{[-R,R]}$ is the indicator of $[-R,R]$;
the magnitude of a radial variable specifies the radius and its sign
selects one of the two opposite phase-space directions. Define the
normalization factor
\begin{equation}
c_R:=|\mathcal D_R|/(2\pi)^M=(R^2/2)^M.
\label{eq:normalization_factor}
\end{equation}
For $\mathbf s\in\mathbb R^M$, the auxiliary direct Monte Carlo
statistic is
\begin{equation}
W_i(\mathbf s):=
\frac{c_R}{n^2}\sum_{a,b=1}^n
\cos\!\left[\mathbf s^{\mathsf T}
(\mathbf x_i^{(a)}-\mathbf y_i^{(b)})\right].
\label{eq:direct_mc_single_setting}
\end{equation}

\begin{lemma}[Unbiasedness of the Estimator]
\label{prop:RB_unbiasedness}
For the measurement protocol and radial distribution defined
above, the estimator satisfies
\begin{equation}
\mathbb E_{\mathsf D}[\widetilde f_R(\rho,\sigma)]
=f_R(\rho,\sigma).
\label{eq:RB_unbiased_target}
\end{equation}
The estimator is defined as
\begin{equation}
\widetilde f_R(\rho,\sigma):=
\frac{1}{N_\theta}\sum_{i=1}^{N_\theta}W_i,
\label{eq:rb_truncated_overlap_estimator}
\end{equation}
where each single-setting statistic $W_i$ is obtained
by analytically averaging its sampled-radial counterpart:
\begin{equation}
W_i:=
\mathbb E_{\mathbf S_i}\!\left[
W_i(\mathbf S_i)\,\middle|\,\mathsf D_i
\right].
\label{eq:rb_radial_average}
\end{equation}
The expectation $\mathbb E_{\mathsf D}$ averages over
both the randomly sampled phase settings and the
corresponding homodyne outcomes.
\end{lemma}

\begin{proof}
By Eq.~\eqref{eq:direct_mc_single_setting}, $|W_i(\mathbf s)|\le c_R$,
so all expectations below are well defined. It suffices to show that
$\mathbb E_{\mathsf D_i}[W_i]=f_R(\rho,\sigma)$ for each setting.
First, fix $\boldsymbol\theta_i$ and $\mathbf s$, and set
$\mathbf u=\boldsymbol\xi(\boldsymbol\theta_i,\mathbf s)$ as in
Eq.~\eqref{eq:slice_map_main}. Taking the real part of
Eq.~\eqref{eq:main_accessible_product} and applying linearity of expectation
to the sum in Eq.~\eqref{eq:direct_mc_single_setting} gives
\begin{equation}
\mathbb E_{\mathsf D_i}\!\left[
W_i(\mathbf s)\,\middle|\,\boldsymbol\theta_i
\right]
=
c_R\operatorname{Re}\!\left[
\chi_\rho(\mathbf u)\chi_\sigma(-\mathbf u)
\right].
\label{eq:direct_mc_conditional_mean}
\end{equation}
The signed polar map in Eq.~\eqref{eq:slice_map_main} has absolute
Jacobian $\prod_{\ell=1}^M|s_\ell|$. Together with the uniform phase
distribution and the radial density in Eq.~\eqref{eq:signed_coordinate_density},
this makes $\mathbf U=\boldsymbol\xi(\boldsymbol\theta_i,\mathbf S_i)$
uniform on $\mathcal D_R$ in Eq.~\eqref{eq:main_disk_domain}.
Averaging Eq.~\eqref{eq:direct_mc_conditional_mean} and using
Eqs.~\eqref{eq:normalization_factor} and~\eqref{eq:main_truncated_target_real}
therefore yields
\begin{equation}
\begin{aligned}
\mathbb E_{\mathsf D_i,\mathbf S_i}\!\left[W_i(\mathbf S_i)\right]
&=
c_R\mathbb E_{\mathbf U}
\operatorname{Re}\!\left[
\chi_\rho(\mathbf U)\chi_\sigma(-\mathbf U)
\right]\\
&=f_R(\rho,\sigma).
\end{aligned}
\label{eq:direct_mc_mean}
\end{equation}
Next, applying the law of iterated expectation to
Eq.~\eqref{eq:rb_radial_average} and using Eq.~\eqref{eq:direct_mc_mean} gives
\begin{equation}
\begin{aligned}
\mathbb E_{\mathsf D_i}[W_i]
&=
\mathbb E_{\mathsf D_i}\!\left[
\mathbb E_{\mathbf S_i}\!\left[
W_i(\mathbf S_i)\,\middle|\,\mathsf D_i
\right]\right]\\
&=
\mathbb E_{\mathsf D_i,\mathbf S_i}\!\left[W_i(\mathbf S_i)\right]
=f_R(\rho,\sigma).
\end{aligned}
\end{equation}

\end{proof}

The conditional expectation in Eq.~\eqref{eq:rb_radial_average} can be
evaluated analytically with $\mathsf D_i$ held fixed.
For one mode, symmetry of the density in Eq.~\eqref{eq:signed_coordinate_density}
gives the real kernel
\begin{equation}
\psi_R(t):=\mathbb E_{S\sim q_R}[e^{iSt}]
=\frac{2}{R^2}\int_0^R s\cos(st)\,ds
=\begin{cases}
\displaystyle\frac{2[Rt\sin(Rt)+\cos(Rt)-1]}{(Rt)^2},&t\ne0,\\[2ex]
1,&t=0.
\end{cases}
\label{eq:rb_kernel}
\end{equation}
Independence of the radial components implies
$\mathbb E_{\mathbf S_i}[\cos(\mathbf S_i^{\mathsf T}\mathbf z)]
=\prod_{\ell=1}^M\psi_R(z_\ell)$ for
$\mathbf z\in\mathbb R^M$. Applying this identity to each pair of
homodyne outcomes gives the explicit single-setting statistic
\begin{equation}
W_i=\frac{c_R}{n^2}\sum_{a=1}^n\sum_{b=1}^n
\prod_{\ell=1}^M
\psi_R\!\left(x_{i,\ell}^{(a)}-y_{i,\ell}^{(b)}\right).
\label{eq:rb_single_setting}
\end{equation}
Thus, analytic radial averaging preserves unbiasedness for $f_R(\rho,\sigma)$ without requiring radial sampling. Moreover, by the law of total variance, the resulting estimator has no greater variance than its sampled-radial counterpart.

Lemma~\ref{prop:RB_unbiasedness} identifies $f_R(\rho,\sigma)$
as the  expectation of the estimator. 
Algorithm~\ref{alg:dipe_cv} summarizes the resulting CV-DIPE protocol.

\begin{algorithm}
\caption{Homodyne estimator for the truncated inner product}
\label{alg:dipe_cv}
\KwInput{phase settings $N_{\theta}$; shots per setting $n$;
truncation radius $R$; modes $M$}
\KwOutput{estimator $\widetilde{f}_{R}(\rho,\sigma)$ of $f_{R}(\rho,\sigma)$}

\For{$i=1,\ldots,N_{\theta}$}{
Sample
$\boldsymbol{\theta}_{i}
\sim\operatorname{Unif}([0,\pi)^{M})$\;

Measure $\rho$ and obtain
$\{\mathbf{x}_{i}^{(a)}\}_{a=1}^{n}$\;

Measure $\sigma$ and obtain
$\{\mathbf{y}_{i}^{(b)}\}_{b=1}^{n}$\;

Compute $W_i$ using
Eq.~\eqref{eq:rb_single_setting}\;
}

\Return{
${\displaystyle
\widetilde{f}_{R}(\rho,\sigma)
=
N_{\theta}^{-1}
\sum_{i=1}^{N_{\theta}}
W_i}$
}
\end{algorithm}

We next establish a variance upper bound that shows the dependence on both $N_\theta$ and $n$. Define
\begin{equation}
v_{R,n}:=
\frac{c_R^2}{n^2}
+c_R\left(1-\frac{1}{n^2}\right).
\label{eq:finite_shot_variance_proxy}
\end{equation}

\begin{theorem}[Variance Upper Bound of the Estimator]
\label{thm:finite_shot_variance}
For fixed $R>0$ and positive integers $n,N_\theta$,
the estimator in
Eq.~\eqref{eq:rb_truncated_overlap_estimator}
satisfies
\begin{equation}
\operatorname{Var}_{\mathsf D}
(\widetilde f_R(\rho,\sigma))
\leq \frac{v_{R,n}}{N_\theta}.
\label{eq:main_finite_shot_variance_bound}
\end{equation}
\end{theorem}

Theorem~\ref{thm:finite_shot_variance} shows that
the variance upper bound scales as $1/N_\theta$, while the
dependence on $n$ is captured by $v_{R,n}$. For $c_R>1$, increasing $n$
tightens the bound. The proof is given in
Appendix~\ref{app:finite_shot_variance}.

\section{SAMPLE COMPLEXITY UNDER DIFFERENT STATE ASSUMPTIONS}
\label{sec:copy_complexity}

Our goal is to estimate $f(\rho,\sigma)$ with error $\epsilon$ and
failure probability at most $\delta$, so that
\begin{equation}
\Pr_{\mathsf D}\!\left(|\widetilde f_R(\rho,\sigma)-f(\rho,\sigma)|\le\epsilon\right)\ge1-\delta.
\label{eq:full_overlap_guarantee}
\end{equation}
Throughout this section, $0<\epsilon,\delta<1$, logarithms are natural,
and $R>0$ and the positive integers $n,N_\theta$ are fixed before sampling.
Probabilities are taken over the complete record $\mathsf D$.

Since the estimator targets $f_R(\rho,\sigma)$ (Lemma~\ref{prop:RB_unbiasedness}), the total error satisfies
\begin{equation}
|\widetilde f_R(\rho,\sigma)-f(\rho,\sigma)|
\le
|\widetilde f_R(\rho,\sigma)-f_R(\rho,\sigma)|
+
|f(\rho,\sigma)-f_R(\rho,\sigma)|.
\label{eq:main_total_error_decomposition}
\end{equation}
An assumption on the states gives a deterministic truncation-bias bound
$|f(\rho,\sigma)-f_R(\rho,\sigma)|\le b(R)$, while concentration controls the statistical error.
We first state a generic finite-sample guarantee
based on Hoeffding's inequality. Each assumption then determines a truncation radius satisfying $b(R)\le\epsilon/2$ and hence a value of
$c_R$, which is substituted into the same guarantee.
This bound uses only $|W_i|\le c_R$ and does not explicitly capture
the dependence on $n$.
We therefore also derive a Bernstein bound in Sec.~\ref{sec:bernstein_bound},
using the variance upper bound $v_{R,n}$ to capture this dependence and potentially
reduce the sufficient setting count.

\subsection{Hoeffding Bound}

Since the single-setting statistics are independent and satisfy
$|W_i|\le c_R$, Hoeffding's inequality~\cite{Hoeffding1963}
gives, for any $t>0$,
\begin{equation}
\Pr_{\mathsf D}\!\left(
|\widetilde f_R(\rho,\sigma)-f_R(\rho,\sigma)|\ge t
\right)
\le
2\exp\!\left(
-\frac{N_\theta t^2}{2c_R^2}
\right).
\label{eq:RB_Hoeffding}
\end{equation}
Together with Eq.~\eqref{eq:main_total_error_decomposition}, this
directly yields the following generic finite-sample guarantee.

\begin{lemma}
\label{lem:generic_copy_complexity}
If $|f(\rho,\sigma)-f_R(\rho,\sigma)|\le b(R)<\epsilon$ and
\begin{equation}
N_\theta
\ge
\frac{2c_R^2\log(2/\delta)}{[\epsilon-b(R)]^2},
\label{eq:generic_setting_count}
\end{equation}
then Eq.~\eqref{eq:full_overlap_guarantee} holds.
In particular, if $b(R)\le\epsilon/2$, it is sufficient to take
\begin{equation}
N_\theta
\ge
\frac{8c_R^2\log(2/\delta)}{\epsilon^2}.
\label{eq:split_error_setting_count}
\end{equation}
\end{lemma}

Lemma~\ref{lem:generic_copy_complexity} gives a bound applicable under either of the two assumptions below. Each assumption gives a computable truncation-bias bound and a truncation radius
satisfying the half-error condition.  The sufficient setting count uses
the boundedness condition $|W_i|\le c_R$, and is therefore independent of $n$.

\paragraph*{\textbf{Mean-Energy Constraint.}}
Let $\hat N=\sum_{\ell=1}^{M}a_\ell^\dagger a_\ell$ be the total
photon-number operator, and let
$\overline E_\tau=\operatorname{Tr}(\tau\hat N)/M$ denote the mean
energy per mode in photon-number units. Suppose
$\overline E_\rho,\overline E_\sigma\le E$.
The truncation-bias bound, proved in
Appendix~\ref{app:proof_energy_copy_complexity}, is
\begin{equation}
\begin{aligned}
|f(\rho,\sigma)-f_R(\rho,\sigma)|
&\le\frac{4ME+2M}{R^2}.
\end{aligned}
\label{eq:energy_bias_certificate}
\end{equation}
Choosing the smallest admissible truncation radius $R^2=\frac{2(4ME+2M)}{\epsilon}$
gives
$c_R=((4ME+2M)/\epsilon)^M$.
Substituting this value of $c_R$ into
Eq.~\eqref{eq:split_error_setting_count} yields the following corollary.

\begin{corollary}[Phase-Setting Complexity under the Mean-Energy Constraint]
\label{thm:energy_copy_complexity}
Under the above mean-energy constraint and truncation radius choice,
Eq.~\eqref{eq:full_overlap_guarantee} holds whenever
\begin{equation}
N_\theta
\ge
\frac{8(4ME+2M)^{2M}}{\epsilon^{2M+2}}
\log\frac{2}{\delta}.
\label{eq:energy_hoeffding_scaling}
\end{equation}
The corresponding total number of state copies consumed
across both platforms is $T_{\mathrm{tot}}=2nN_\theta$.
\end{corollary}

For fixed $M$ and $E$, the scaling in Eq.~\eqref{eq:energy_hoeffding_scaling} reflects two distinct sources of statistical cost. The factor $\epsilon^{-2}$ is the usual concentration cost associated with estimating the truncated inner product to additive accuracy $O(\epsilon)$, whereas the additional factor $\epsilon^{-2M}$ arises from truncation. Under only a mean-energy constraint, the truncation-bias bound decays algebraically as $R^{-2}$, so the above truncation radius ensures $|f(\rho,\sigma)-f_R(\rho,\sigma)|\le \epsilon/2$ with $R^2=O(\epsilon^{-1})$. By Eq.~\eqref{eq:normalization_factor}, this translates into $c_R^2=O(\epsilon^{-2M})$. Thus, the $\epsilon^{-2M-2}$ dependence is not caused by concentration alone, but by the combination of algebraic tail control and the $2M$-dimensional phase-space volume. The bound is therefore conservative but fully state-independent within the prescribed mean-energy class.

\paragraph*{\textbf{Gaussian States.}}
Suppose in addition that $\rho$ and $\sigma$ are Gaussian states,
subject to the same known mean-energy bound.
The resulting truncation-bias bound, proved in
Appendix~\ref{app:gaussian_tail_certificate}, is
\begin{equation}
|f(\rho,\sigma)-f_R(\rho,\sigma)|
\le
M\exp\!\left(-\frac{R^2}{2(4ME+2M)}\right).
\label{eq:physical_gaussian_tail_constants}
\end{equation}
The smallest admissible truncation radius is
$R^2=2(4ME+2M)\log(2M/\epsilon)$, for which
$c_R=[(4ME+2M)\log(2M/\epsilon)]^M$.
Substituting this value of $c_R$ into
Eq.~\eqref{eq:split_error_setting_count}  gives the Gaussian counterpart.

\begin{corollary}[Phase-Setting Complexity for Gaussian States]
\label{thm:gaussian_tail_copy_complexity}
Under the above assumptions on the Gaussian states and the stated choice of truncation radius,
Eq.~\eqref{eq:full_overlap_guarantee} holds whenever
\begin{equation}
N_\theta
\ge
\frac{8}{\epsilon^2}
\left[(4ME+2M)\log\frac{2M}{\epsilon}\right]^{2M}
\log\frac{2}{\delta}.
\label{eq:gaussian_hoeffding_scaling}
\end{equation}
\end{corollary}

The stronger Gaussian tails improve the accuracy scaling at fixed $M$ and $E$ [Eqs.~\eqref{eq:energy_hoeffding_scaling} and~\eqref{eq:gaussian_hoeffding_scaling}]. The Gaussian truncation-bias bound decays exponentially in $R^2$, so the above truncation radius ensures $|f(\rho,\sigma)-f_R(\rho,\sigma)|\le\epsilon/2$ with $R^2=O(\log(1/\epsilon))$, rather than $O(\epsilon^{-1})$. Consequently, the truncation-induced contribution to the normalization factor becomes polylogarithmic in $1/\epsilon$, and the dominant accuracy dependence of the sufficient setting count is reduced from $\epsilon^{-2M-2}$ to $\epsilon^{-2}[\log(2M/\epsilon)]^{2M}$. This shows explicitly how additional structural information about the state class can substantially reduce the cost of controlling the unbounded phase-space tail. The dependence on $M$, however, remains nontrivial through both the power $2M$ and the energy-dependent factor $(4ME+2M)^{2M}$.

Taken together, Corollaries~\ref{thm:energy_copy_complexity} and~\ref{thm:gaussian_tail_copy_complexity} separate the effect of statistical concentration from that of the tail bounds obtained under the two assumptions. Both results use the same Hoeffding bound and therefore have the same $\epsilon^{-2}\log(1/\delta)$ dependence from statistical concentration. Their different accuracy scalings arise from the algebraic and exponential decay of the respective truncation-bias bounds. In this sense, the assumptions on the states affect the sample complexity through phase-space truncation rather than through the estimator itself.

The Hoeffding bound uses only $|W_i|\le c_R$, so the sufficient setting count is independent of $n$. Consequently, the dependence on finite-shot sampling is not reflected in Corollaries~\ref{thm:energy_copy_complexity} and~\ref{thm:gaussian_tail_copy_complexity}. To complement the Hoeffding bound, we therefore also derive a Bernstein bound that incorporates both this boundedness condition and $v_{R,n}$. In contrast to the Hoeffding bound, the Bernstein bound explicitly captures the dependence on $n$ through the single-setting variance upper bound.

\subsection{Bernstein Bound}
\label{sec:bernstein_bound}
Combining the variance upper bound in Theorem~\ref{thm:finite_shot_variance}
with $|W_i|\le c_R$, the standard Bernstein
inequality~\cite{Boucheron2013Concentration} gives a complementary
finite-sample guarantee for any prescribed $n$.

\begin{proposition}
\label{prop:bernstein_concentration}
For any $t>0$,
\begin{equation}
\Pr_{\mathsf D}\!\left(|\widetilde f_R(\rho,\sigma)-f_R(\rho,\sigma)|\ge t\right)
\le2\exp\!\left[-\frac{N_\theta t^2}
{2v_{R,n}+\frac43c_Rt}\right].
\label{eq:bernstein_tail}
\end{equation}
If $|f(\rho,\sigma)-f_R(\rho,\sigma)|\le b(R)<\epsilon$, then Eq.~\eqref{eq:full_overlap_guarantee}
holds whenever
\begin{equation}
N_\theta\ge
\left[\frac{2v_{R,n}}{[\epsilon-b(R)]^2}
+\frac{4c_R}{3[\epsilon-b(R)]}\right]\log(2/\delta).
\label{eq:bernstein_setting_count}
\end{equation}
In particular, if $b(R)\le\epsilon/2$, it is sufficient that
\begin{equation}
N_\theta\ge
\left[\frac{8v_{R,n}}{\epsilon^2}
+\frac{8c_R}{3\epsilon}\right]\log(2/\delta).
\label{eq:bernstein_half_budget_count}
\end{equation}
\end{proposition}

The Bernstein bound in Eq.~\eqref{eq:bernstein_half_budget_count} replaces the squared normalization factor $c_R^2$ by
$v_{R,n}$ and adds a term linear in $c_R$.
Comparing Eqs.~\eqref{eq:split_error_setting_count} and
\eqref{eq:bernstein_half_budget_count} at the same $R,n,\epsilon,\delta$,
the Bernstein threshold is smaller  when
\begin{equation}
v_{R,n}+\frac{\epsilon c_R}{3}<c_R^2.
\label{eq:bernstein_hoeffding_comparison}
\end{equation}
For $c_R>1$, the condition for the Bernstein bound to be tighter than the Hoeffding bound is
$(c_R-1)(1-n^{-2})>\frac{\epsilon}{3}$.
Thus, the Hoeffding bound is tighter at $n=1$, whereas the Bernstein bound becomes tighter when the reduction in the variance upper bound outweighs the contribution of its additional linear term.

The same truncation radius choices used in
Corollaries~\ref{thm:energy_copy_complexity} and
\ref{thm:gaussian_tail_copy_complexity} satisfy
$|f(\rho,\sigma)-f_R(\rho,\sigma)|\le\epsilon/2$ under their respective assumptions. Substituting their values of $c_R$
into Eq.~\eqref{eq:bernstein_half_budget_count} gives, for the
mean-energy constraint,
\begin{equation}
N_\theta\ge\frac{8}{\epsilon^2}\log\frac{2}{\delta}
\Biggl[\frac{1}{n^2}\left(\frac{4ME+2M}{\epsilon}\right)^{2M}
+\left(1-\frac{1}{n^2}+\frac{\epsilon}{3}\right)
\left(\frac{4ME+2M}{\epsilon}\right)^M\Biggr],
\label{eq:energy_copy_complexity_scaling}
\end{equation}
and, for Gaussian states satisfying the same mean-energy constraint,
\begin{equation}
N_\theta\ge\frac{8}{\epsilon^2}\log\frac{2}{\delta}\Biggl[
\frac{1}{n^2}[(4ME+2M)\log(2M/\epsilon)]^{2M}
+\left(1-\frac{1}{n^2}+\frac{\epsilon}{3}\right)
[(4ME+2M)\log(2M/\epsilon)]^M\Biggr].
\label{eq:gaussian_tail_copy_scaling}
\end{equation}
Eqs.~\eqref{eq:energy_copy_complexity_scaling}
and~\eqref{eq:gaussian_tail_copy_scaling} give sufficient phase-setting
counts for general energy-constrained states and Gaussian states satisfying the same constraint, respectively, for any
prescribed $n$. Both bounds contain a contribution proportional
to $c_R^2/n^2$ together with terms linear in $c_R$. For $c_R>1$,
increasing $n$ therefore lowers the sufficient setting count, making the
tradeoff between shots per setting and independent phase settings explicit.
The Gaussian-state assumption replaces the truncation-induced powers of $1/\epsilon$
in the mean-energy bound by powers of $\log(2M/\epsilon)$.

Together, the Hoeffding and Bernstein bounds provide sufficient
measurement resources for estimating the full inner product at a prescribed
accuracy and confidence. The assumption on the states determines a truncation radius that
controls the bias, while concentration determines a sufficient number of
phase settings to control statistical fluctuations.
In Sec.~\ref{sec:numerics_plan}, we use numerical simulations across representative CV state families to investigate how these two error sources jointly determine the estimator's finite-sample mean-squared error (MSE).

\section{Numerical Experiments}
\label{sec:numerics_plan}
Using representative CV state families with analytically known inner products, we evaluate three aspects of the estimator's finite-sample performance: its dependence on \(R\), the mode- and energy-dependent MSE after pointwise truncation radius selection, and its performance across different exact inner-product values using a fixed family-level truncation radius.

Note that the rigorous bounds derived in Sec.~\ref{sec:copy_complexity} give choices of the truncation radius under the stated assumptions, but they need not identify the truncation radius that minimizes the actual finite-sample MSE for a particular state pair. Indeed, the exact truncation bias and estimator variance depend on the detailed phase-space structure of the states. 

\begin{figure}[t]
\centering
\includegraphics[width=0.55\linewidth]{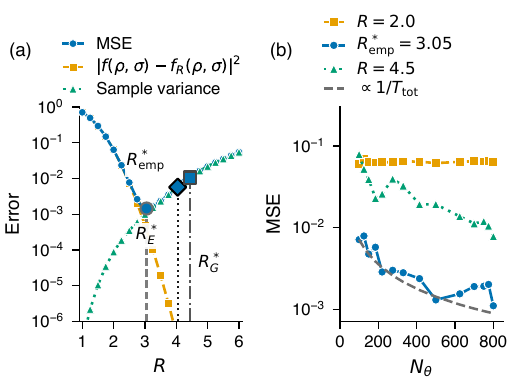}
\caption{
Truncation radius dependence for a same-state coherent-product benchmark
\(|\alpha\rangle^{\otimes 2}\), with \(\alpha=0.6+0.2i\),
\(M=2\) modes, mean energy per mode
\(\bar E=|\alpha|^2=0.4\), and \(n=10\) shots per state per setting.
(a) The MSE is evaluated as the analytic \(|f(\rho,\sigma)-f_R(\rho,\sigma)|^2\) plus the sample
variance estimating \(\operatorname{Var}(\widetilde f_R(\rho,\sigma))\), as in
Eq.~\eqref{eq:numerical_mse_definition}, at \(N_\theta=800\) shared phase
settings. The truncation radius is selected using 80 repetitions, and the curves use
80 independent validation repetitions on the truncation radius grid fixed before validation.
The square, circle, and diamond mark \(R_G^*=4.44\),
\(R_{\rm emp}^*=3.05\), and \(R_E^*=4.04\), respectively.
$R_G^*$ and $R_E^*$ are defined in
Eqs.~\eqref{eq:Gaussian_cutoff_guide} and~\eqref{eq:energy_cutoff_guide}.
(b) The direct empirical MSE in
Eq.~\eqref{eq:numerical_empirical_mse} is shown versus the number of shared
phase settings \(N_\theta\), using 30 repetitions per point at
\(R=2.0\), \(R_{\rm emp}^*=3.05\), and \(R=4.5\).
The gray dashed reference line indicates
$T_{\mathrm{tot}}^{-1}$ scaling.
}
\label{fig:truncation_tradeoff}
\end{figure}

Since the estimator is unbiased for the
truncated inner product, the MSE is
\begin{equation}
\operatorname{MSE}_{f(\rho,\sigma)}(R)
=
\mathbb{E}_{\mathsf D}\!\left[(\widetilde f_R(\rho,\sigma)-f(\rho,\sigma))^2\right]
=
|f(\rho,\sigma)-f_R(\rho,\sigma)|^2+\operatorname{Var}(\widetilde f_R(\rho,\sigma)).
\label{eq:numerical_mse_definition}
\end{equation}
Thus, the MSE consists of the squared truncation bias and the estimator variance.
Under the mean-energy constraint of Sec.~\ref{sec:copy_complexity}, the truncation-bias bound in Eq.~\eqref{eq:energy_bias_certificate} together with the finite-shot variance upper bound
in Eq.~\eqref{eq:main_finite_shot_variance_bound} gives
\begin{equation}
\operatorname{MSE}_{f(\rho,\sigma)}(R)
\le
\frac{(4ME+2M)^2}{R^4}
+
\frac{v_{R,n}}{N_\theta}.
\label{eq:energy_mse_upper_bound}
\end{equation}
Minimizing this MSE upper bound defines the truncation radius
\begin{equation}
R_E^*
:=
\arg\min_{R>0}
\left[\frac{(4ME+2M)^2}{R^4}+\frac{v_{R,n}}{N_\theta}\right].
\label{eq:energy_cutoff_guide}
\end{equation}
Similarly, for Gaussian states satisfying the same mean-energy constraint,
the Gaussian truncation-bias bound
in Eq.~\eqref{eq:physical_gaussian_tail_constants} gives
\begin{equation}
\operatorname{MSE}_{f(\rho,\sigma)}(R)
\le
M^2\exp\!\left[-\frac{R^2}{4ME+2M}\right]
+
\frac{v_{R,n}}{N_\theta}.
\label{eq:gaussian_mse_upper_bound}
\end{equation}
For Gaussian states, the truncation radius selected by minimizing this MSE upper bound is
\begin{equation}
R_G^*
:=
\arg\min_{R>0}
\left[M^2e^{-\frac{R^2}{4ME+2M}}+\frac{v_{R,n}}{N_\theta}\right].
\label{eq:Gaussian_cutoff_guide}
\end{equation}
Because the bounds in Eqs.~\eqref{eq:energy_mse_upper_bound}
and~\eqref{eq:gaussian_mse_upper_bound} are uniform upper bounds within the respective state classes, the corresponding
\(R_E^*\) and \(R_G^*\) need not coincide with the truncation radius that
minimizes the actual finite-sample MSE for a particular state pair.
When the MSE is evaluated directly from \(N_{\rm rep}\) independent numerical
repetitions, we use
\begin{equation}
\operatorname{MSE}_{f(\rho,\sigma)}(R)
=
\frac{1}{N_{\rm rep}}
\sum_{k=1}^{N_{\rm rep}}
\left(\widetilde f_R^{(k)}(\rho,\sigma)-f(\rho,\sigma)\right)^2.
\label{eq:numerical_empirical_mse}
\end{equation}
For pointwise selection based on
Eq.~\eqref{eq:numerical_empirical_mse}, we denote the empirical
MSE-minimizing truncation radius by
\begin{equation}
R_{\rm emp}^*
:=
\arg\min_{R>0}
\operatorname{MSE}_{f(\rho,\sigma)}(R).
\label{eq:empirical_cutoff}
\end{equation}
For the coherent-state benchmark in
Fig.~\ref{fig:truncation_tradeoff}(a), we instead minimize the analytic
\(|f(\rho,\sigma)-f_R(\rho,\sigma)|^2\) plus the sample variance over independent selection repetitions;
\(R_{\rm emp}^*\) denotes the resulting truncation radius.
Eq.~\eqref{eq:numerical_mse_definition} characterizes the
bias--variance tradeoff, while
Eqs.~\eqref{eq:energy_cutoff_guide} and
\eqref{eq:Gaussian_cutoff_guide} define $R_E^*$ and $R_G^*$, respectively, and
Eq.~\eqref{eq:empirical_cutoff} defines the empirical MSE-minimizing truncation radius
for the direct-MSE benchmarks.
\subsection{Truncation Radius Dependence}

We first investigate the effect of the truncation radius on finite-sample estimation, using the two-mode coherent-product state $|\alpha\rangle^{\otimes 2}$ as a benchmark. Fig.~\ref{fig:truncation_tradeoff}(a) illustrates the resulting bias–variance tradeoff. At each $R$, the estimator variance is estimated from 80 independent numerical repetitions, and the MSE is computed as the sum of this empirical variance and the exact squared truncation bias. This comparison identifies how the choice of truncation radius affects the estimator's finite-sample accuracy.

Fig.~\ref{fig:truncation_tradeoff}(a) shows opposite trends in the two error
components as the truncation radius increases. The squared truncation bias falls
from order unity to below $10^{-6}$, whereas the empirical variance increases
from below $10^{-6}$ to above $10^{-2}$. Consequently, the MSE is
nonmonotonic. The selection data give $R_{\rm emp}^*=3.05$, at which the
independently evaluated MSE decomposition is approximately
$1.4\times10^{-3}$. The two radii, $R_E^*=4.04$ and $R_G^*=4.44$, defined in Eqs.~\eqref{eq:energy_cutoff_guide} and
\eqref{eq:Gaussian_cutoff_guide}, both lie on the variance-dominated branch.
These choices of radius are conservative because they minimize
the uniform upper bounds for the respective state classes in
Eqs.~\eqref{eq:energy_mse_upper_bound} and~\eqref{eq:gaussian_mse_upper_bound},
rather than the actual MSE of the present coherent-state benchmark. The
resulting conservative tail control favors larger truncation radii, where the exact
truncation bias is already small but the estimator variance is correspondingly
larger.

Fig.~\ref{fig:truncation_tradeoff}(b) compares the dependence of the MSE on $N_\theta$ at three fixed truncation radii. For $R=2.0$, the MSE remains near $6\times10^{-2}$ as
$N_\theta$ increases from 100 to 800. For $R=4.5$, it decreases from
approximately $8\times10^{-2}$ to $8\times10^{-3}$, but remains above the
result obtained at $R_{\rm emp}^*$. At $R_{\rm emp}^*=3.05$, the MSE
decreases from approximately $7\times10^{-3}$ to $1.1\times10^{-3}$ and
approximately follows the displayed $T_{\mathrm{tot}}^{-1}$ reference over the tested range of $N_\theta$.

\begin{figure}[t]
\centering
\includegraphics[width=0.60\linewidth]{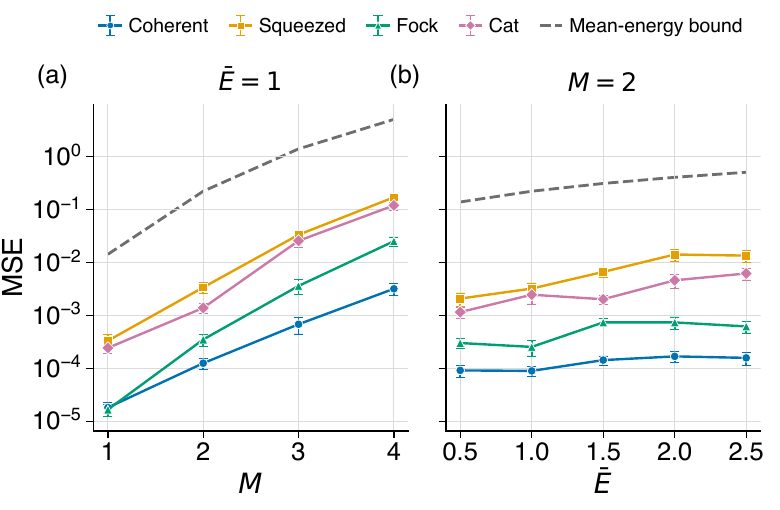}
\caption{%
Finite-sample MSE versus mode number and mean energy per mode for same-state
comparisons with $f(\rho,\sigma)=1$.
(a) MSE versus the mode number $M=1,2,3,4$ at fixed mean energy per
mode $\bar E=1$.
(b) MSE versus $\bar E=0.5,1.0,1.5,2.0,2.5$ at fixed $M=2$.
The benchmarks are coherent states $\bigotimes_{\ell=1}^{M}|\alpha_\ell\rangle$
with $|\alpha_\ell|^2=\bar E$;
squeezed-vacuum states $[S(\xi)|0\rangle]^{\otimes M}$ with $\xi\ge0$ and $\sinh^2\xi=\bar E$;
Fock states $\bigotimes_{\ell=1}^{M}|m_\ell\rangle$ with
$\sum_\ell m_\ell=M\bar E$ and integer occupations differing by at most one;
and even-cat states
$\bigotimes_{\ell=1}^{M}[\mathcal N_\ell(|\beta_\ell\rangle+|-\beta_\ell\rangle)]$
with $|\beta_\ell|^2\tanh(|\beta_\ell|^2)=\bar E$.
Here, $S(\xi):=\exp[(\xi^*\hat a^2-\xi\hat a^{\dagger 2})/2]$ is the
single-mode squeezing operator, and
$\mathcal N_\ell=[2(1+e^{-2|\beta_\ell|^2})]^{-1/2}$ is the normalization
coefficient of the single-mode even-cat state;
the coherent and cat amplitudes have mode phases $2\pi(\ell-1)/M$.
Error bars denote one standard error of the MSE.
The gray dashed curve shows the mean-energy MSE upper bound
in Eq.~\eqref{eq:energy_mse_upper_bound}, evaluated at $R_E^*$
defined in Eq.~\eqref{eq:energy_cutoff_guide}.
All results use $N_\theta=5000$ shared phase settings and $n=20$
shots per state per setting.
}
\label{fig:mode_energy_scaling}
\end{figure}

At fixed $R$, increasing $N_\theta$ reduces the estimator variance but does
not change the truncation bias. Thus, the small truncation radius $R=2.0$ is
bias-dominated and gains little from additional phase settings, whereas the
larger truncation radius $R=4.5$ suppresses the truncation bias at the cost of a larger estimator variance
that requires more settings to reduce. The intermediate choice
$R_{\rm emp}^*=3.05$ balances these effects, keeping the truncation bias below the
dominant statistical error while allowing the estimator variance to decrease
approximately as $N_\theta^{-1}$ over the tested range. Since
$T_{\mathrm{tot}}$ is proportional to $N_\theta$ at fixed $n$, this yields the observed approximate
$T_{\mathrm{tot}}^{-1}$ MSE scaling.

\subsection{MSE versus Mode Number and Energy}

We next examine how the finite-sample error changes with system size and
mean energy per mode for different state families. For each state--parameter point
shown in Fig.~\ref{fig:mode_energy_scaling}, $R_{\rm emp}^*$ is selected separately
according to Eq.~\eqref{eq:empirical_cutoff}.
The MSE as a function of $R$ is estimated from 20 independent selection
repetitions, after which the selected truncation radius is fixed and evaluated using
30 independent validation repetitions. Each repetition uses
$N_\theta=5000$ shared phase settings and $n=20$ shots per state per
setting. The error bars are obtained from the squared errors across the
validation repetitions, while the gray dashed curve shows the mean-energy MSE
upper bound in Eq.~\eqref{eq:energy_mse_upper_bound}, evaluated at $R_E^*$.

Fig.~\ref{fig:mode_energy_scaling}(a) shows that the MSE increases with the number
of modes for all four state families. At $M=1$, the coherent and Fock
results are on the $10^{-5}$ scale, whereas the squeezed and even-cat
results are on the $10^{-4}$ scale. At $M=4$, the coherent-state MSE is
approximately $3.2\times10^{-3}$, the Fock-state MSE is on the
$10^{-2}$ scale, and the squeezed and even-cat MSEs reach the
$10^{-1}$ scale. The mean-energy MSE upper bound remains above all empirical
results over the tested mode range.

At fixed $M=2$, Fig.~\ref{fig:mode_energy_scaling}(b) shows a clear
state-family-dependent energy response. The coherent-state MSE remains near
$10^{-4}$ throughout $\bar E=0.5$--$2.5$. Over the same interval, the
squeezed-state MSE increases from approximately $2\times10^{-3}$ to
$1.4\times10^{-2}$, while the even-cat MSE increases from approximately
$1.2\times10^{-3}$ to $6.2\times10^{-3}$. The Fock-state MSE remains
below $10^{-3}$ and varies more weakly, with a slight decrease at the
largest energy. The mean-energy MSE upper bound again remains above the
empirical results throughout the displayed range.

\begin{figure}[t]
\centering
\includegraphics[width=0.50\linewidth,trim=8bp 0bp 8bp 0bp,clip]{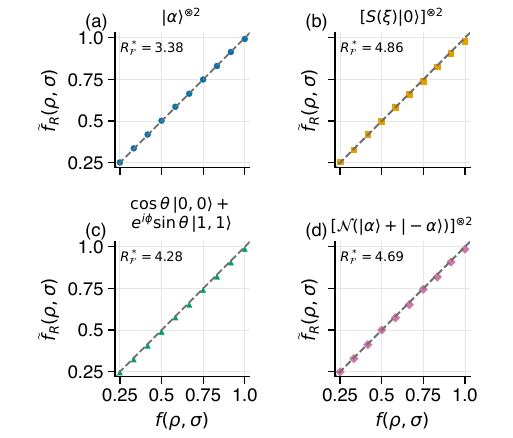}
\caption{%
Estimated versus exact inner products for four two-mode benchmark
families, with mean energy per mode $\bar E=0.5$,
$N_\theta=8000$ shared phase settings, and $n=20$ shots per state per
setting. Panels show (a) coherent states, (b) squeezed-vacuum states,
(c) the Fock-sector superposition
$\cos\theta\,|0,0\rangle+e^{i\phi}\sin\theta\,|1,1\rangle$, and
(d) even-cat product states.
Here, $S(\xi)$ is the single-mode squeezing operator defined in the caption of
Fig.~\ref{fig:mode_energy_scaling}, and
$\mathcal N_+=[2(1+e^{-2|\alpha|^2})]^{-1/2}$
normalizes the single-mode even-cat state
$\mathcal N_+(|\alpha\rangle+|-\alpha\rangle)$.
In panel (c), $\theta$ and $\phi$ denote the state-mixing angle and
relative phase, respectively, rather than phase settings.
A single family-level truncation radius $R_{\mathcal F}^*$ is used for all ten target inner-product values
within each panel, with values $3.38$, $4.86$, $4.28$, and $4.69$
for panels (a)--(d), respectively.
The horizontal axis gives the exact inner product $f(\rho,\sigma)$, while the vertical axis
gives the mean of 30 independent validation estimates
$\widetilde f_R(\rho,\sigma)$. The dashed line is the identity line.
}
\label{fig:overlap_calibration}
\end{figure}

The two trends have different origins. The dependence on $M$ is primarily
dimensional: increasing the number of modes enlarges the sampled phase space
and amplifies the statistical cost through $c_R$.
The dependence on energy is instead state-family specific, because increasing
energy modifies different phase-space features in different ways. In
particular, displacement changes the phase but not the Gaussian envelope of
a coherent-state characteristic function, whereas squeezing and cat-state
separation modify the relevant phase-space width and structure. Thus, the
mode dependence reflects mainly the dimensional cost of the estimator, while
the energy dependence probes the detailed phase-space structure of the state
family. The state-independent bound in Eq.~\eqref{eq:energy_mse_upper_bound}
evaluated at $R_E^*$ does not exploit this
additional structure and is therefore conservative relative to the empirical MSE.

\subsection{Estimation across Inner-Product Values}

Finally, we examine the estimator away from the same-state case $f(\rho,\sigma)=1$.
For each state family $\mathcal F$ in Fig.~\ref{fig:overlap_calibration}, a
single family-level truncation radius is selected rather than a separate radius
for each target inner-product value. The truncation radius search is driven by the family-averaged
empirical MSE
\begin{equation}
\label{eq:family_empirical_mse}
\frac{1}{10\times20}
\sum_{j=1}^{10}\sum_{k=1}^{20}
\left[\widetilde f_R^{(k)}(\rho_j,\sigma_j)-f(\rho_j,\sigma_j)\right]^2,
\end{equation}
where $(\rho_j,\sigma_j)$ denotes the $j$th state pair, $f(\rho_j,\sigma_j)$ is its exact inner product, and the average
is taken over ten target inner-product values and 20 independent selection repetitions.
The search first identifies truncation radii with small values of
Eq.~\eqref{eq:family_empirical_mse}, after which a small set of candidates is compared
using 50 independent confirmation repetitions to determine the final
$R_{\mathcal F}^*$. Thus, $R_{\mathcal F}^*$ is selected empirically from
family-level performance rather than from the MSE upper bounds used to obtain $R_E^*$ and $R_G^*$. Once confirmed, $R_{\mathcal F}^*$ is fixed before validation and used for all
ten target inner-product values in that family. The plotted values are then obtained
from a further 30 independent validation repetitions, which do not enter
either the selection or confirmation stages.

Fig.~\ref{fig:overlap_calibration} shows that the mean estimates remain close
to the identity line over the tested range $f(\rho,\sigma)\in[0.25,1]$ for all four
state families. The residual deviations occur on both sides of the
reference line and remain small compared with the full variation of the
inner product. Hence, a single family-level truncation radius continues to provide accurate
inner-product estimation as the underlying state pair is varied within each
benchmark family.

This provides a more stringent test than selecting a separate truncation radius for
each state pair. For fixed $R$, the estimator is unbiased for
$f_R(\rho,\sigma)$, so the error relative to $f(\rho,\sigma)$
is governed by the truncation bias and
finite-sampling fluctuations around $f_R(\rho,\sigma)$. Crucially, the truncation bias
depends on the phase-space tail of the specific state pair and is not
determined by $f(\rho,\sigma)$ alone. Using a common $R_{\mathcal F}^*$
across ten different state pairs therefore tests whether a single
family-level truncation radius remains adequate throughout the state family,
rather than whether the truncation radius can be retuned to each individual target.
The near-diagonal behavior indicates that, over the tested range, the
family-dependent variation of the relevant phase-space tails remains
sufficiently controlled that one family-level truncation radius can keep the truncation
bias small. In this sense, $R_{\mathcal F}^*$
can be reused without target-specific retuning.

Taken together, the numerical experiments characterize three complementary
finite-sample properties of the protocol.
Fig.~\ref{fig:truncation_tradeoff} reveals how the bias--variance
tradeoff depends on the truncation radius and identifies the range of $T_{\mathrm{tot}}$ over which the approximate $T_{\mathrm{tot}}^{-1}$ Monte
Carlo MSE scaling becomes visible. Fig.~\ref{fig:mode_energy_scaling} shows
how the residual error depends on mode number and on the phase-space
structure of different state families. Finally,
Fig.~\ref{fig:overlap_calibration} shows that a single empirically selected
family-level truncation radius can remain effective across multiple state pairs
without target-specific retuning.

\section{Conclusion}
\label{sec:conclusions}

We have developed a homodyne-only protocol for CV-DIPE between independently prepared multimode quantum states. By combining product-of-disks truncation with analytic radial averaging, the protocol yields a bounded estimator that is unbiased for the truncated inner product. We establish finite-sample guarantees under a mean-energy constraint, with improved truncation-bias bounds for Gaussian states.

The truncation radius $R$ governs the tradeoff between truncation bias and statistical error. Although prior-dependent bias bounds provide sufficient radius choices, they can be conservative and need not minimize the finite-sample MSE. Our numerical results demonstrate that smaller, empirically selected radii can yield better bias–variance balance. Developing reliable data-driven radius selection methods and extending the analysis to experimental noise and drift are important directions for future work.

\begin{acknowledgments}
This work was supported by the National Natural Science Foundation of China
(Grant No.~62471126), the Frontier Technologies R\&D Program of Jiangsu
(Grant No.~BF2025066), the Jiangsu Key R\&D Program Project
(Grant No.~BE2023011-2), and the Fundamental Research Funds for the Central
Universities (Grant No.~2242022k60001).
\end{acknowledgments}

\bibliographystyle{apsrev4-2-threeauthors}
\bibliography{main}

\clearpage

\appendix
\appendixcontents

\section{Phase-Space Notation and Preliminary Identities}
\label{app:proof_preliminaries}

We summarize here the notation and identities used repeatedly in the
subsequent proofs. The purpose of this appendix is to connect the statistics of homodyne measurements at shared phase settings with the phase-space inner-product integrand and
to express the truncated phase-space integral as an expectation over the
truncation domain.

We use the conventions of the main text,
\[
\hat{\mathbf{R}}
=
(\hat q_1,\hat p_1,\ldots,\hat q_M,\hat p_M)^{\mathsf T},
\qquad
W(\mathbf{u})=e^{i\mathbf{u}^{\mathsf T}\hat{\mathbf{R}}},
\]
with $[\hat R_j,\hat R_k]=i\Omega_{jk}I$ and
$\Omega=\bigoplus_{\ell=1}^{M}
\begin{psmallmatrix}0&1\\-1&0\end{psmallmatrix}$, and
\[
\chi_\tau(\mathbf{u})
=
\operatorname{Tr}[\tau W(\mathbf{u})].
\]
For the two states \(\rho\) and \(\sigma\), let
$f(\rho,\sigma)=\operatorname{Tr}(\rho\sigma)$ denote the full inner product.
For truncation radius $R>0$, the truncated inner product is
\begin{equation}
f_R(\rho,\sigma)
=
\frac{1}{(2\pi)^M}
\int_{\mathcal D_R}
\chi_\rho(\mathbf u)\chi_\sigma(-\mathbf u)\,d^{2M}\mathbf{u},
\label{eq:app_truncated_overlap}
\end{equation}
where $
\mathcal D_R
=
\prod_{\ell=1}^{M}
\left\{
(u_{q,\ell},u_{p,\ell})\in\mathbb R^2:
u_{q,\ell}^2+u_{p,\ell}^2\le R^2
\right\}$
is the truncation domain, a modewise product of disks.
The normalization factor is
\[
c_R
=
\frac{|\mathcal D_R|}{(2\pi)^M}
=
\left(\frac{R^2}{2}\right)^M.
\]
For $n$ shots per state per setting, the single-setting variance upper bound is
$v_{R,n}=c_R^2/n^2+c_R(1-1/n^2)$, as in
Eq.~\eqref{eq:finite_shot_variance_proxy}.
For
\(\boldsymbol{\theta}=(\theta_1,\ldots,\theta_M)\in[0,\pi)^M\) and
\(\mathbf{s}=(s_1,\ldots,s_M)^{\mathsf T}\), define $\boldsymbol{\xi}(\boldsymbol{\theta},\mathbf{s})
=
\big(
s_1\cos\theta_1,
s_1\sin\theta_1,
\ldots,
s_M\cos\theta_M,
s_M\sin\theta_M
\big)^{\mathsf T}.$
For each mode, the parameter \(s_\ell\) is sampled over the interval
\([-R,R]\). Its magnitude \(|s_\ell|\) determines the radial distance,
while the two possible signs correspond to the two opposite directions
associated with a fixed phase \(\theta_\ell\).

The first identity required below is the Fourier-slice relation. Let
\(p_\tau(\mathbf{x}\mid\boldsymbol{\theta})\) denote the joint probability
density of the \(M\)-mode homodyne outcomes for state \(\tau\) at phase
\(\boldsymbol{\theta}\). For
\(\mathbf{x}\sim p_\tau(\cdot\mid\boldsymbol{\theta})\),
\begin{equation}
\mathbb E_{\mathbf{x}}
\!\left[
e^{i\mathbf{s}^{\mathsf T}\mathbf{x}}
\,\middle|\,
\boldsymbol{\theta}
\right]
=
\chi_\tau
\!\left(
\boldsymbol{\xi}(\boldsymbol{\theta},\mathbf{s})
\right).
\label{eq:app_fourier_slice_identity}
\end{equation}
Hence, for conditionally independent homodyne outcomes at a shared phase setting
\(\mathbf{x}\sim p_\rho(\cdot\mid\boldsymbol{\theta})\) and
\(\mathbf{y}\sim p_\sigma(\cdot\mid\boldsymbol{\theta})\),
\begin{equation}
\mathbb E_{\mathbf{x},\mathbf{y}}
\!\left[
e^{i\mathbf{s}^{\mathsf T}(\mathbf{x}-\mathbf{y})}
\,\middle|\,
\boldsymbol{\theta}
\right]
=
\chi_\rho\!\left(\boldsymbol\xi(\boldsymbol\theta,\mathbf s)\right)
\chi_\sigma\!\left(-\boldsymbol\xi(\boldsymbol\theta,\mathbf s)\right).
\label{eq:app_matched_fourier_slice}
\end{equation}
Thus, data from homodyne measurements at shared phase settings enable estimation of the characteristic-function product at specified phase-space points. To recover \(f_R(\rho,\sigma)\), these pointwise values must then be combined so as to reproduce the phase-space integral over \(\mathcal D_R\). We therefore choose the phase and signed-radial sampling law such that the induced phase-space point is uniformly distributed over \(\mathcal D_R\). Let
\[
\boldsymbol{\theta}
\sim
\operatorname{Unif}([0,\pi)^M),
\qquad
\mathbf{S}=(S_1,\ldots,S_M)^{\mathsf T},
\]
where the components \(S_\ell\) are independent with density
\begin{equation}
q_R(s)
=
\frac{|s|}{R^2}
\mathrm{1}_{[-R,R]}(s).
\label{eq:app_signed_coordinate_density}
\end{equation}
The vector $\mathbf S$ is sampled independently of $\boldsymbol\theta$.
For each mode, the joint density of \((\theta_\ell,S_\ell)\) is
\(|s_\ell|/(\pi R^2)\), while the change of variables
\((u_{q,\ell},u_{p,\ell})
=
(s_\ell\cos\theta_\ell,s_\ell\sin\theta_\ell)\)
has Jacobian
\(du_{q,\ell}du_{p,\ell}
=
|s_\ell|\,ds_\ell d\theta_\ell\).
Hence the induced density in each disk is \(1/(\pi R^2)\).
Independence across modes therefore gives
\begin{equation}
\mathbf{U}
=
\boldsymbol{\xi}(\boldsymbol{\theta},\mathbf{S})
\sim
\operatorname{Unif}(\mathcal D_R).
\label{eq:app_uniform_disk_vector}
\end{equation}
Consequently, for any integrable function \(k\),
\[
\mathbb E_{\mathbf{U}}[k(\mathbf{U})]
=
\frac{1}{|\mathcal D_R|}
\int_{\mathcal D_R}k(\mathbf{u})\,d^{2M}\mathbf{u}.
\]
Applying this identity to
$k(\mathbf u)=\chi_\rho(\mathbf u)\chi_\sigma(-\mathbf u)$ and using
\(c_R\), Eq.~\eqref{eq:app_truncated_overlap} becomes
\[
f_R(\rho,\sigma)
=
c_R
\mathbb E_{\mathbf U}
\left[\chi_\rho(\mathbf U)\chi_\sigma(-\mathbf U)\right].
\]
Moreover, $\chi_\tau(-\mathbf u)=\overline{\chi_\tau(\mathbf u)}$.
Since $\mathcal D_R$ is invariant under $\mathbf u\mapsto-\mathbf u$,
the imaginary part of $\chi_\rho(\mathbf u)\chi_\sigma(-\mathbf u)$
integrates to zero.
Therefore,
\begin{equation}
f_R(\rho,\sigma)
=
c_R
\mathbb E_{\mathbf U}
\operatorname{Re}\left[\chi_\rho(\mathbf U)\chi_\sigma(-\mathbf U)\right].
\label{eq:app_disk_expectation_identity}
\end{equation}
Eqs.~\eqref{eq:app_matched_fourier_slice} and
\eqref{eq:app_disk_expectation_identity} provide the two basic relations
needed for the estimator analysis. The former connects homodyne outcomes at shared phase settings with the phase-space integrand, while the latter expresses the
truncated inner product as an expectation over the phase settings and radial
sampling variables. These relations also underlie the finite-shot variance analysis in Appendix~\ref{app:finite_shot_variance}.

\section{ variance of the Estimator}
\label{app:proof_rb_unbiasedness_variance}
\label{app:finite_shot_variance}

Here we prove the finite-shot variance upper bound in
Theorem~\ref{thm:finite_shot_variance}.
We quantify the estimator variance
and retain the explicit dependence on the number \(n\) of homodyne shots per
phase setting. The derivation first evaluates the second moments of the empirical
characteristic functions, then averages the resulting bound over the
truncation domain using the Plancherel identity, and
finally uses independence across the \(N_\theta\) phase settings.

For $i=1,\ldots,N_\theta$, let $\boldsymbol\theta_i$ and
$\mathbf S_i$ follow the independent phase and radial laws in
Appendix~\ref{app:proof_preliminaries}. Conditional on
$\boldsymbol\theta_i$, the vectors
$\mathbf X_i^{(a)}\sim p_\rho(\cdot\mid\boldsymbol\theta_i)$ and
$\mathbf Y_i^{(b)}\sim p_\sigma(\cdot\mid\boldsymbol\theta_i)$,
for $a,b=1,\ldots,n$, are independent shots within each platform,
and the two platform samples are conditionally independent.
Write $\mathbf X_i^{(1:n)}:=(\mathbf X_i^{(1)},\ldots,\mathbf X_i^{(n)})$
and similarly for $\mathbf Y_i^{(1:n)}$.
The records $\mathsf D_i:=(\boldsymbol\theta_i,
\mathbf X_i^{(1:n)},\mathbf Y_i^{(1:n)})$ are independent across
settings, $\mathsf D:=(\mathsf D_1,\ldots,\mathsf D_{N_\theta})$,
and $\mathbf S_i$ is independent of $\mathsf D_i$.
Fix $\boldsymbol\theta_i=\boldsymbol\theta$ and
$\mathbf S_i=\mathbf s$, and write
$\mathbf u=\boldsymbol\xi(\boldsymbol\theta,\mathbf s)$.
The empirical characteristic functions are
\begin{equation}
\begin{aligned}
\widetilde\chi_{\rho,i}(\mathbf u)&:=\frac{1}{n}\sum_{a=1}^{n}
e^{i\mathbf{s}^{\mathsf T}\mathbf{X}_i^{(a)}},\\
\widetilde\chi_{\sigma,i}(\mathbf u)&:=\frac{1}{n}\sum_{b=1}^{n}
e^{i\mathbf{s}^{\mathsf T}\mathbf{Y}_i^{(b)}}.
\end{aligned}
\label{eq:app_empirical_characteristic_functions}
\end{equation}
Expanding the first estimate gives
\begin{align}
\mathbb E_{\mathbf{X}_i^{(1:n)}}[|\widetilde\chi_{\rho,i}(\mathbf u)|^2\mid\boldsymbol{\theta},\mathbf{s}]
&=
\frac1{n^2}
\sum_{a,a'}
\mathbb E_{\mathbf{X}_i^{(a)},\mathbf{X}_i^{(a')}}
\left[
 e^{i\mathbf{s}^{\mathsf T}
 (\mathbf{X}_i^{(a)}-\mathbf{X}_i^{(a')})}
 \middle|
 \boldsymbol{\theta},\mathbf{s}
\right].
\end{align}
There are \(n\) diagonal terms, each equal to one, and \(n(n-1)\)
off-diagonal terms. For \(a\neq a'\), independence of distinct shots and the
Fourier-slice identity give
\begin{equation}
\mathbb E_{\mathbf{X}_i^{(a)},\mathbf{X}_i^{(a')}}
\left[
 e^{i\mathbf{s}^{\mathsf T}
 (\mathbf{X}_i^{(a)}-\mathbf{X}_i^{(a')})}
 \middle|
 \boldsymbol{\theta},\mathbf{s}
\right]
=
|\chi_\rho(\mathbf{u})|^2.
\end{equation}
Therefore,
\begin{equation}
\mathbb E_{\mathbf{X}_i^{(1:n)}}[|\widetilde\chi_{\rho,i}(\mathbf u)|^2\mid\boldsymbol{\theta},\mathbf{s}]
=
\frac1n+
\left(1-\frac1n\right)|\chi_\rho(\mathbf{u})|^2.
\label{eq:app_A_second_moment_final}
\end{equation}
Similarly,
\begin{equation}
\mathbb E_{\mathbf{Y}_i^{(1:n)}}[|\widetilde\chi_{\sigma,i}(\mathbf u)|^2\mid\boldsymbol{\theta},\mathbf{s}]
=
\frac1n+
\left(1-\frac1n\right)|\chi_\sigma(\mathbf{u})|^2.
\label{eq:app_B_second_moment_final}
\end{equation}
Because the two platform records are conditionally independent and
\((\operatorname{Re}z)^2\le|z|^2\),
\begin{align}
&\mathbb E_{\mathbf{X}_i^{(1:n)},\mathbf{Y}_i^{(1:n)}}
\left[
\left(c_R\operatorname{Re}\!\left[\widetilde\chi_{\rho,i}(\mathbf u)\overline{\widetilde\chi_{\sigma,i}(\mathbf u)}\right]\right)^2
\middle|
\boldsymbol{\theta},\mathbf{s}
\right]
\nonumber\\
&\quad\le
c_R^2
\left[
\frac1n+
\left(1-\frac1n\right)|\chi_\rho(\mathbf{u})|^2
\right]
\left[
\frac1n+
\left(1-\frac1n\right)|\chi_\sigma(\mathbf{u})|^2
\right].
\label{eq:app_direct_second_moment_final}
\end{align}
Use the uniform random vector $\mathbf U$ defined in
Eq.~\eqref{eq:app_uniform_disk_vector}.
Under the characteristic-function convention of the main text, the
Plancherel identity gives
\begin{equation}
\frac1{(2\pi)^M}
\int_{\mathbb R^{2M}}
|\chi_\tau(\mathbf{u})|^2\,d^{2M}\mathbf{u}
=
\operatorname{Tr}(\tau^2)
\le1.
\end{equation}
Since \(\mathbf{U}\) is uniform on \(\mathcal D_R\),
\begin{equation}
c_R\mathbb E_{\mathbf{U}}[|\chi_\rho(\mathbf{U})|^2]\le1,
\qquad
c_R\mathbb E_{\mathbf{U}}[|\chi_\sigma(\mathbf{U})|^2]\le1.
\label{eq:app_single_plancherel_bounds_final}
\end{equation}
Moreover, \(|\chi_\tau(\mathbf{u})|\le1\) implies
\begin{equation}
c_R\mathbb E_{\mathbf{U}}[|\chi_\rho(\mathbf{U})|^2|\chi_\sigma(\mathbf{U})|^2]\le1.
\label{eq:app_product_plancherel_bound_final}
\end{equation}
With $v_{R,n}$ as specified in Appendix~\ref{app:proof_preliminaries},
expanding Eq.~\eqref{eq:app_direct_second_moment_final}, averaging over \(\mathbf{U}\),
and applying Eqs.~\eqref{eq:app_single_plancherel_bounds_final} and
\eqref{eq:app_product_plancherel_bound_final}, we obtain
\begin{align}
\mathbb E_{\mathsf D_i,\mathbf{S}_i}[\left(c_R\operatorname{Re}\!\left[\widetilde\chi_{\rho,i}(\mathbf U)\overline{\widetilde\chi_{\sigma,i}(\mathbf U)}\right]\right)^2]
&\le
\frac{c_R^2}{n^2}
+
c_R\left[
\frac{2}{n}\left(1-\frac1n\right)
+
\left(1-\frac1n\right)^2
\right]
\nonumber\\
&=
\frac{c_R^2}{n^2}
+
c_R\left(1-\frac1{n^2}\right)=v_{R,n}.
\label{eq:app_direct_second_moment_bound_final}
\end{align}
The radial integral in Eq.~\eqref{eq:rb_radial_average} can be
written for these records as
\begin{equation}
W_i=\mathbb E_{\mathbf S_i}\!\left[
c_R\operatorname{Re}\!\left[
\widetilde\chi_{\rho,i}(\mathbf U)
\overline{\widetilde\chi_{\sigma,i}(\mathbf U)}\right]
\,\middle|\,\mathsf D_i\right].
\label{eq:app_radial_integral_for_variance}
\end{equation}
Applying Jensen's inequality to this integral and averaging
over $\mathsf D_i$ gives $\mathbb E_{\mathsf D_i}[W_i^2]\le v_{R,n}$
by Eq.~\eqref{eq:app_direct_second_moment_bound_final}.
Using $\operatorname{Var}_{\mathsf D_i}(W_i)\le
\mathbb E_{\mathsf D_i}[W_i^2]$, we obtain
\begin{equation}
\operatorname{Var}_{\mathsf D_i}(W_i)
\le
v_{R,n}.
\label{eq:app_finite_shot_setting_variance_final}
\end{equation}
Independence of the records across settings implies independence of the
single-setting statistics $W_i$. Hence, the estimator variance satisfies
\begin{equation}  
\operatorname{Var}_{\mathsf D}(\widetilde f_R(\rho,\sigma))
=
\frac1{N_\theta^2}
\sum_{i=1}^{N_\theta}
\operatorname{Var}_{\mathsf D_i}(W_i)
\le
\frac{v_{R,n}}{N_\theta}.
\label{eq:app_estimator_finite_shot_variance_final}
\end{equation}
This proves Theorem~\ref{thm:finite_shot_variance}.
For \(c_R\ge1\), the variance upper bound can be written as
\[
v_{R,n}
=
c_R+\frac{c_R^2-c_R}{n^2},
\]
and is therefore nonincreasing in \(n\). It equals \(c_R^2\) at \(n=1\)
and approaches \(c_R\) as \(n\) increases. Accordingly, the estimator
variance upper bound starts at \(c_R^2/N_\theta\) and converges to
\(c_R/N_\theta\). This limiting value refers only to the upper bound and
does not constitute a lower bound on the actual variance. For arbitrary \(R>0\), the boundedness
\(|W_i|\le c_R\) also implies
\[
\operatorname{Var}_{\mathsf D_i}(W_i)\le c_R^2.
\]
Hence
\(\min\{c_R^2,v_{R,n}\}\)
is always a valid single-setting variance upper bound. When \(c_R<1\), the
range-based bound \(c_R^2\) is no larger than \(v_{R,n}\).

\section{Truncation-Bias Bounds}
\label{app:proof_copy_complexity}

Using the inner-product representation in Eq.~\eqref{eq:phase_space_identity_main}
and the definition of $f_R(\rho,\sigma)$ in
Eq.~\eqref{eq:app_truncated_overlap}, the truncation bias is determined
entirely by the phase-space region omitted by the truncation:
\begin{equation}
|f(\rho,\sigma)-f_R(\rho,\sigma)|
=
\left|
\frac{1}{(2\pi)^M}
\int_{\mathbb R^{2M}\setminus\mathcal D_R}
\chi_\rho(\mathbf u)\chi_\sigma(-\mathbf u)\,d^{2M}\mathbf u
\right|
\le
\frac{1}{(2\pi)^M}
\int_{\mathbb R^{2M}\setminus\mathcal D_R}
|\chi_\rho(\mathbf u)|
|\chi_\sigma(-\mathbf u)|
\,d^{2M}\mathbf u .
\label{eq:app_truncation_bias_tail}
\end{equation}

Thus, controlling the truncation bias reduces to controlling the
phase-space tail of the characteristic functions outside
$\mathcal D_R$. The available bound depends on the prior information
about the states. Under only a mean-energy constraint, the energy controls
a weighted second moment of the characteristic functions, which yields an
algebraically decaying truncation-bias bound. For Gaussian states, the characteristic
functions have Gaussian envelopes, and the same energy constraint gives bounds on their covariance matrices, leading to an exponentially decaying truncation-bias bound.

We derive these two truncation-bias bounds below. Substituting the
resulting bounds and the corresponding truncation radius choices into
Lemma~\ref{lem:generic_copy_complexity} gives
Corollaries~\ref{thm:energy_copy_complexity} and~\ref{thm:gaussian_tail_copy_complexity}.

\subsection{Mean-Energy Truncation-Bias Bounds}
\label{app:proof_energy_copy_complexity}

We prove Eq.~\eqref{eq:energy_bias_certificate} by bounding a
weighted second moment of the characteristic function under the
mean-energy constraint.

\paragraph*{\textbf{Energy Controls a Weighted Phase-Space Moment.}}
Let
\begin{equation}
\hat N=\sum_{\ell=1}^{M}a_\ell^\dagger a_\ell,
\qquad
\overline E_\tau
:=
\frac1M\operatorname{Tr}(\tau\hat N),
\label{eq:app_mean_energy_definition}
\end{equation}
where $\hat N$ is the total photon-number operator and
$\overline E_\tau$ is the mean energy per mode in photon-number units.  Consider
\begin{equation}
I_\tau
:=
\frac{1}{(2\pi)^M}
\int_{\mathbb R^{2M}}
\|\mathbf{u}\|_2^2|\chi_\tau(\mathbf{u})|^2\,d^{2M}\mathbf{u}.
\label{eq:app_energy_weighted_moment}
\end{equation}
We first take $\tau$ to have finite photon-number support. Under the
characteristic-function convention of the main text, the Plancherel identity and
$[\hat R_j,W(\mathbf{u})]=-(\Omega \mathbf{u})_jW(\mathbf{u})$ imply
\begin{equation}
I_\tau
=
\sum_{j=1}^{2M}
\|[\hat R_j,\tau]\|_2^2.
\label{eq:app_energy_commutator_identity}
\end{equation}
For each quadrature operator,
\begin{equation}
\|[\hat R_j,\tau]\|_2^2
=
2\operatorname{Tr}(\hat R_j^2\tau^2)
-2\operatorname{Tr}(\hat R_j\tau\hat R_j\tau).
\label{eq:app_commutator_norm_expansion}
\end{equation}
The trace in the second term is nonnegative because
\begin{equation}
\operatorname{Tr}(\hat R_j\tau\hat R_j\tau)
=
\|\sqrt\tau\hat R_j\sqrt\tau\|_2^2\ge0.
\label{eq:app_nonnegative_quadrature_trace}
\end{equation}
Using
\begin{equation}
\sum_{j=1}^{2M}\hat R_j^2=2\hat N+M
\label{eq:app_quadrature_energy_identity}
\end{equation}
and $0\le\tau^2\le\tau$, we obtain
\begin{equation}
\begin{aligned}
I_\tau
&=
4\operatorname{Tr}(\tau^2\hat N)
+2M\operatorname{Tr}(\tau^2)
-2\sum_{j=1}^{2M}
\|\sqrt\tau\hat R_j\sqrt\tau\|_2^2
\\
&\le
4M\overline E_\tau+2M.
\end{aligned}
\label{eq:app_energy_moment_bound}
\end{equation}
For a general finite-energy state, let $P_K$ project onto total photon number
at most $K$ and define
\begin{equation}
\tau_K
:=
\frac{P_K\tau P_K}{\operatorname{Tr}(P_K\tau)}.
\label{eq:app_finite_support_approximation}
\end{equation}
Then $\tau_K\to\tau$ in trace norm and
$\operatorname{Tr}(\tau_K\hat N)\to M\overline E_\tau$. Since
$\|W(\mathbf{u})\|=1$, the characteristic functions converge pointwise. Fatou's
lemma therefore extends Eq.~\eqref{eq:app_energy_moment_bound} to every
finite-energy state:
\begin{equation}
I_\tau
\le
4M\overline E_\tau+2M.
\label{eq:app_energy_moment_bound_general}
\end{equation}

\paragraph*{\textbf{From the Energy Moment to the Truncation Bias.}}
If $\mathbf{u}\notin\mathcal D_R$, at least one mode has radius larger than $R$,
and therefore $\|\mathbf{u}\|_2\ge R$. Using Eq.~\eqref{eq:app_truncation_bias_tail}, we obtain
\begin{equation}
\begin{aligned}
|f(\rho,\sigma)-f_R(\rho,\sigma)|
&\le
\frac{1}{(2\pi)^M}
\int_{\mathbb R^{2M}\setminus\mathcal D_R}
|\chi_\rho(\mathbf{u})|\,|\chi_\sigma(-\mathbf{u})|\,d^{2M}\mathbf{u}
\\
&\le
\frac{1}{R^2}
\frac{1}{(2\pi)^M}
\int_{\mathbb R^{2M}}
\|\mathbf{u}\|_2^2
|\chi_\rho(\mathbf{u})|\,|\chi_\sigma(-\mathbf{u})|\,d^{2M}\mathbf{u}.
\end{aligned}
\label{eq:app_energy_tail_first_step}
\end{equation}
Applying Cauchy--Schwarz to the last integral and using
Eq.~\eqref{eq:app_energy_moment_bound_general} gives
\begin{equation}
|f(\rho,\sigma)-f_R(\rho,\sigma)|
\le
\frac{\sqrt{I_\rho I_\sigma}}{R^2}
\le
\frac{
\sqrt{
(4M\overline E_\rho+2M)
(4M\overline E_\sigma+2M)
}
}{R^2}.
\label{eq:app_energy_bias_certificate}
\end{equation}
If
\begin{equation}
\overline E_\rho,\overline E_\sigma\le E,
\label{eq:app_mean_energy_assumption}
\end{equation}
then the truncation-bias bound simplifies to
\begin{equation}
|f(\rho,\sigma)-f_R(\rho,\sigma)|
\le
\frac{4ME+2M}{R^2}.
\label{eq:app_energy_uniform_bias_certificate}
\end{equation}
The inverse-square dependence has a simple interpretation: finite energy
controls a quadratic phase-space moment, and outside the truncation domain the
weight $\|\mathbf{u}\|_2^2$ is at least $R^2$.

\subsection{Gaussian Truncation-Bias Bounds}
\label{app:proof_gaussian_copy_complexity}
\label{app:gaussian_tail_certificate}

We now derive the truncation-bias bound for Gaussian states.
The main point is that Gaussianity fixes the functional form of the
characteristic-function envelope, while the mean-energy constraint controls
how broad this envelope can be. The argument proceeds in three steps.
First, the energy bound and the uncertainty relation provide a uniform
lower bound on the covariance matrix. Second, the modulus of the product
of the two characteristic functions is normalized as an auxiliary
classical Gaussian density. Finally, the probability that this auxiliary
Gaussian vector leaves the truncation domain is bounded mode by mode.

\paragraph{\textbf{Energy Bounds on the Covariance Matrix.}}
Let $\tau\in\{\rho,\sigma\}$ satisfy $\overline E_\tau\le E$,
with the mean energy per mode defined in Eq.~\eqref{eq:app_mean_energy_definition}.
Define the phase-space displacement vector
$\mathbf d_\tau:=\operatorname{Tr}(\tau\hat{\mathbf R})$. The centered quadratures are
$\Delta\hat R_j:=\hat R_j-d_{\tau,j}I$.
We define the covariance matrix by
\begin{equation}
(\Gamma_\tau)_{jk}
:=
\operatorname{Tr}
\!\left[
\tau\{
\Delta\hat R_j,\Delta\hat R_k
\}
\right],
\label{eq:app_covariance_definition}
\end{equation}
where $\{A,B\}:=AB+BA$ denotes the anticommutator. 
With this convention, $\Gamma_\tau$ is twice the usual centered
quadrature covariance matrix:
$(\Gamma_\tau)_{jj}=2\operatorname{Var}_\tau(\hat R_j)$.
Hence, the quadratic form
$\mathbf v^{\mathsf T}\Gamma_\tau\mathbf v/2$
is the variance of the quadrature
$\mathbf v^{\mathsf T}\hat{\mathbf R}$.
In particular, the vacuum state has
$\Gamma_\tau=I_{2M}$.
Using
Eq.~\eqref{eq:app_quadrature_energy_identity},
we obtain
\begin{equation}
\begin{aligned}
\operatorname{Tr}\Gamma_\tau
&=
2\sum_{j=1}^{2M}
\left[
\operatorname{Tr}(\tau\hat R_j^2)-d_{\tau,j}^2
\right]
\\
&=
4\operatorname{Tr}(\tau\hat N)
+2M
-2\|\mathbf d_\tau\|_2^2
\\
&\le 4ME+2M.
\end{aligned}
\label{eq:app_energy_covariance_trace}
\end{equation}
Thus, the mean-energy constraint bounds the total quadrature fluctuation.
By itself, however, this trace bound does not prevent one covariance
eigenvalue from becoming arbitrarily small. A lower bound in every
phase-space direction follows from the uncertainty relation.
To see this explicitly, let $\mathbf v\in\mathbb R^{2M}$ be any unit
vector and define
$\mathbf w:=-\Omega\mathbf v$,
$Q_{\mathbf v}:=\mathbf v^{\mathsf T}\hat{\mathbf R}$, and
$P_{\mathbf v}:=\mathbf w^{\mathsf T}\hat{\mathbf R}$.
Since $\Omega^{\mathsf T}\Omega=I_{2M}$ and $\Omega^2=-I_{2M}$,
$\mathbf w$ is also a unit vector, and
\begin{equation}
[Q_{\mathbf v},P_{\mathbf v}]
=
i\mathbf v^{\mathsf T}\Omega\mathbf w\,I
=
iI.
\label{eq:app_conjugate_commutator}
\end{equation}
Thus, $Q_{\mathbf v}$ and $P_{\mathbf v}$ form a canonically conjugate
pair. The Heisenberg uncertainty relation gives
\begin{equation}
\operatorname{Var}_\tau(Q_{\mathbf v})
\operatorname{Var}_\tau(P_{\mathbf v})
\ge \frac14.
\label{eq:app_quadrature_uncertainty}
\end{equation}
Using
$\operatorname{Var}_\tau(Q_{\mathbf v})
=\mathbf v^{\mathsf T}\Gamma_\tau\mathbf v/2$
and the analogous expression for $P_{\mathbf v}$ in
Eq.~\eqref{eq:app_quadrature_uncertainty}, we obtain
\begin{equation}
\bigl(
\mathbf v^{\mathsf T}\Gamma_\tau\mathbf v
\bigr)
\bigl(
\mathbf w^{\mathsf T}\Gamma_\tau\mathbf w
\bigr)
\ge 1.
\label{eq:app_energy_covariance_uncertainty}
\end{equation}
Because $\Gamma_\tau\succeq0$, $\mathbf w$ is a unit vector, and
$\operatorname{Tr}\Gamma_\tau\le 4ME+2M$,
$\mathbf w^{\mathsf T}\Gamma_\tau\mathbf w \le 4ME+2M$.
Therefore,
$\mathbf v^{\mathsf T}\Gamma_\tau\mathbf v\ge (4ME+2M)^{-1}$
for every unit vector $\mathbf v$, which is equivalent to
\begin{equation}
\Gamma_\tau
\succeq
(4ME+2M)^{-1}I_{2M}.
\label{eq:app_energy_covariance_lower_bound}
\end{equation}
Here $A\succeq B$ means that $A-B$ is positive semidefinite.
Eq.~\eqref{eq:app_energy_covariance_lower_bound} expresses the combined effect of the energy
constraint and the uncertainty relation: the energy bound prevents
arbitrarily large total quadrature fluctuations, while the uncertainty
relation prevents the fluctuation in any conjugate direction from
collapsing arbitrarily close to zero.

For the two states, define
$G:=\Gamma_\rho+\Gamma_\sigma$.
This matrix determines the decay envelope of the product of their
characteristic functions. Applying Eq.~\eqref{eq:app_energy_covariance_lower_bound} to both states gives
\begin{equation}
G\succeq\frac{2}{4ME+2M}I_{2M},
\qquad
G^{-1}\preceq\frac{4ME+2M}{2}I_{2M}.
\label{eq:app_energy_G_inverse_bound}
\end{equation}
The second relation gives an upper bound on $G^{-1}$ in terms of the mean energy.

For later use, let $P_\ell:\mathbb R^{2M}\rightarrow\mathbb R^2$
extract the two phase-space coordinates of mode $\ell$, so that
$P_\ell P_\ell^{\mathsf T}=I_2$, and define
$\Sigma_\ell:=P_\ell G^{-1}P_\ell^{\mathsf T}$.
The matrix $2\Sigma_\ell$ will be the covariance matrix of the
$\ell$th two-dimensional block of an auxiliary Gaussian variable
introduced below. Eq.~\eqref{eq:app_energy_G_inverse_bound} implies
$2\Sigma_\ell\preceq (4ME+2M)I_2$.
Define the state-dependent parameter
$\mu_G:=\max_\ell\lambda_{\max}(\Sigma_\ell)$,
the largest eigenvalue among the matrices $\Sigma_\ell$.
Consequently,
\begin{equation}
\mu_G\le\frac{4ME+2M}{2}.
\label{eq:app_gaussian_mu}
\end{equation}

\paragraph{\textbf{Gaussian Envelope and Its Normalization.}}
Gaussianity now enters explicitly. For a Gaussian state,
\begin{equation}
\chi_\tau(\mathbf u)
=
\exp\!\left[
-\frac14\mathbf u^{\mathsf T}\Gamma_\tau\mathbf u
+i\mathbf d_\tau^{\mathsf T}\mathbf u
\right].
\label{eq:app_gaussian_characteristic_function}
\end{equation}
The displacement vector contributes only a phase to the characteristic
function and therefore does not affect its magnitude. Hence,
\begin{equation}
|\chi_\rho(\mathbf u)\chi_\sigma(-\mathbf u)|
=
\exp\!\left(
-\frac14\mathbf u^{\mathsf T}G\mathbf u
\right).
\label{eq:app_gaussian_envelope}
\end{equation}
Thus, the covariance matrices determine the decay of the phase-space
tail, whereas the relative displacement affects only the oscillatory
phase of the inner-product integrand.
The function in Eq.~\eqref{eq:app_gaussian_envelope} has a Gaussian form but is not yet a normalized
probability density. We therefore introduce the normalization factor
\begin{equation}
\kappa_G
:=
\frac{1}{(2\pi)^M}
\int_{\mathbb R^{2M}}
e^{-\mathbf u^{\mathsf T}G\mathbf u/4}
\,d^{2M}\mathbf u.
\label{eq:app_gaussian_mass_definition}
\end{equation}
Evaluating the Gaussian integral gives
\begin{equation}
\kappa_G
=
\frac{2^M}{\sqrt{\det G}}.
\label{eq:app_gaussian_normalization}
\end{equation}
The normalization factor $\kappa_G$ generally differs from the inner product because the latter also contains the relative displacement phase.
Moreover, the Cauchy--Schwarz inequality and the Plancherel identity give
\begin{equation}
\begin{aligned}
\kappa_G
&=
\frac{1}{(2\pi)^M}
\int_{\mathbb R^{2M}}
|\chi_\rho(\mathbf u)|
|\chi_\sigma(-\mathbf u)|
\,d^{2M}\mathbf u
\\
&\le
\sqrt{
\operatorname{Tr}(\rho^2)
\operatorname{Tr}(\sigma^2)
}
\le 1.
\end{aligned}
\label{eq:app_gaussian_mass_bound}
\end{equation}
This normalization bound is important because it allows the tail integral
to be converted into a probability without introducing an uncontrolled
state-dependent prefactor.
Consider the normalized density
\begin{equation}
\frac{1}{(2\pi)^M\kappa_G}
e^{-\mathbf z^{\mathsf T}G\mathbf z/4}.
\label{eq:app_gaussian_density_definition}
\end{equation}
By Eq.~\eqref{eq:app_gaussian_mass_definition}, this density integrates to one, and comparison with the standard
multivariate Gaussian density shows that it is the probability density
of the auxiliary classical random vector
\begin{equation}
\mathbf Z\sim\mathcal N(\mathbf 0,2G^{-1}).
\label{eq:app_gaussian_auxiliary_vector}
\end{equation}
This auxiliary variable should not be confused with the physical
quadrature distribution of either quantum state; it is introduced only
to represent the normalized characteristic-function envelope.
Using the phase-space tail representation of the truncation bias
in Eq.~\eqref{eq:app_truncation_bias_tail},
Eq.~\eqref{eq:app_gaussian_envelope} therefore gives
\begin{equation}
\begin{aligned}
|f(\rho,\sigma)-f_R(\rho,\sigma)|
&\le
\frac{1}{(2\pi)^M}
\int_{\mathbb R^{2M}\setminus\mathcal D_R}
e^{-\mathbf u^{\mathsf T}G\mathbf u/4}
\,d^{2M}\mathbf u
\\
&=
\kappa_G\,
\Pr_{\mathbf Z}
\!\left(
\mathbf Z\notin\mathcal D_R
\right).
\end{aligned}
\label{eq:app_gaussian_bias_probability}
\end{equation}
The quantum tail-integral problem has thus been reduced to a classical
Gaussian tail-probability problem.

\paragraph{\textbf{Energy-Controlled Modewise Tails.}}
Because $\mathcal D_R$ is a product of $M$ two-dimensional disks, it is
natural to control the probability of leaving the truncation domain
mode by mode. For mode $\ell$,
\begin{equation}
P_\ell\mathbf Z
\sim
\mathcal N(\mathbf 0,2\Sigma_\ell).
\label{eq:app_gaussian_mode_marginal}
\end{equation}
Equivalently,
$P_\ell\mathbf Z \overset{d}{=} \sqrt{2}\,\Sigma_\ell^{1/2}\mathbf g$,
where $\mathbf g\sim\mathcal N(\mathbf 0,I_2)$ and
$\overset{d}{=}$ denotes equality in distribution.
Since $2\Sigma_\ell\preceq (4ME+2M)I_2$,
\begin{equation}
\left\|
\sqrt{2}\,\Sigma_\ell^{1/2}\mathbf g
\right\|_2^2
\le
(4ME+2M)\|\mathbf g\|_2^2.
\label{eq:app_gaussian_block_norm_bound}
\end{equation}
Therefore,
\begin{equation}
\Pr_{\mathbf Z}
\!\left(
\|P_\ell\mathbf Z\|_2>R
\right)
\le
\Pr_{\mathbf g}
\!\left(
\|\mathbf g\|_2>\frac{R}{\sqrt{4ME+2M}}
\right).
\label{eq:app_gaussian_block_tail_comparison}
\end{equation}

For a two-dimensional standard Gaussian vector, direct integration in
polar coordinates gives, for $a\ge0$,
\begin{equation}
\Pr_{\mathbf g}(\|\mathbf g\|_2>a)
=
\frac{1}{2\pi}
\int_0^{2\pi}
\int_a^\infty
e^{-r^2/2}r\,dr\,d\phi
=
e^{-a^2/2}.
\label{eq:app_gaussian_two_dimensional_tail}
\end{equation}
Combining Eqs.~\eqref{eq:app_gaussian_block_tail_comparison}
and~\eqref{eq:app_gaussian_two_dimensional_tail}, each mode block satisfies
\begin{equation}
\Pr_{\mathbf Z}
\!\left(
\|P_\ell\mathbf Z\|_2>R
\right)
\le
e^{-\frac{R^2}{2(4ME+2M)}}.
\label{eq:app_gaussian_block_tail}
\end{equation}
The complement of the truncation domain is the union of the
events that at least one mode block has radius larger than $R$:
$\{\mathbf Z\notin\mathcal D_R\} = \bigcup_{\ell=1}^M \{\|P_\ell\mathbf Z\|_2>R\}$.
The union bound, which does not require independence between the mode
blocks, therefore gives
\begin{equation}
\Pr_{\mathbf Z}
\!\left(
\mathbf Z\notin\mathcal D_R
\right)
\le
\sum_{\ell=1}^M
\Pr_{\mathbf Z}
\!\left(
\|P_\ell\mathbf Z\|_2>R
\right)
\le
M e^{-\frac{R^2}{2(4ME+2M)}}.
\label{eq:app_gaussian_union_bound}
\end{equation}
Combining Eqs.~\eqref{eq:app_gaussian_mass_bound},
\eqref{eq:app_gaussian_bias_probability}, and~\eqref{eq:app_gaussian_union_bound}, we obtain the state-independent
Gaussian truncation-bias bound
\begin{equation}
|f(\rho,\sigma)-f_R(\rho,\sigma)|
\le
M e^{-\frac{R^2}{2(4ME+2M)}}.
\label{eq:app_gaussian_bias_certificate}
\end{equation}

If the actual covariance information is available, the same argument
can retain $\mu_G$ instead of replacing it by
the energy-only bound in Eq.~\eqref{eq:app_gaussian_mu}. Since
$2\Sigma_\ell\preceq2\mu_G I_2$, this gives the sharper,
state-dependent truncation-bias bound
\begin{equation}
|f(\rho,\sigma)-f_R(\rho,\sigma)|
\le
\kappa_G M
e^{-\frac{R^2}{4\mu_G}}.
\label{eq:app_gaussian_state_dependent_certificate}
\end{equation}
Eq.~\eqref{eq:app_gaussian_bias_certificate} is the uniform truncation-bias bound requiring only the Gaussian
assumption and the known mean-energy bound, whereas Eq.~\eqref{eq:app_gaussian_state_dependent_certificate} shows how
additional covariance information can sharpen the tail control. Finally, choosing
$R^2=2(4ME+2M)\log(2M/\epsilon)$
ensures $|f(\rho,\sigma)-f_R(\rho,\sigma)|\le\epsilon/2$. For this choice,
$c_R=[(4ME+2M)\log(2M/\epsilon)]^M$.
Substituting this value into Eq.~\eqref{eq:split_error_setting_count} of
Lemma~\ref{lem:generic_copy_complexity} yields
Corollary~\ref{thm:gaussian_tail_copy_complexity}.

\end{document}